\documentclass[onecolumn,secnumarabic,amssymb, nobibnotes, aps, prd]{revtex4}
\usepackage{amsmath} \usepackage{amssymb}
\usepackage{graphicx} 
\usepackage{epstopdf}
\usepackage{amsmath}
\usepackage{amsmath,amssymb,amsthm,amsfonts,mathrsfs,bm,verbatim}
\usepackage{graphicx,subfigure}
\usepackage{appendix}

\newcommand{\bea}{\begin{eqnarray}}
\newcommand{\eea}{\end{eqnarray}}

\begin{document}

\title{Thermodynamic geometric analysis of 3D charged black holes under f(R) gravity }%

\author{Wen-Xiang Chen}
\affiliation{Department of Astronomy, School of Physics and Materials Science, Guangzhou University, Guangzhou 510006, China}
\email{wxchen4277@qq.com}
\author{Yao-Guang Zheng}
\affiliation{Department of Physics, College of Sciences, Northeastern University, Shenyang 110819, China}

\begin{abstract}
This article investigates 3D charged black holes within the scope of f(R) gravity, focusing on their thermodynamic attributes. The research primarily examines minor fluctuations around these black holes' equilibrium states and delves into their modified thermodynamic entropy. Utilizing geometric thermodynamics (GTD), the study evaluates the curvature scalar's role in pinpointing phase transition points in these black holes. A key finding is that several 3D charged black holes under f(R) gravity display thermodynamic properties akin to an ideal gas when their initial curvature scalar remains constant. Conversely, with a non-constant curvature scalar and a cosmological constant term that includes a negative exponent, these black holes exhibit characteristics similar to a van der Waals gas. The article outlines general solutions for scenarios involving non-negative powers and specific solutions for cases with negative powers. Notably, under certain conditions, a phase transition resembling that of a van der Waals gas is observed, suggesting a strong correlation between the black hole's fate and the cosmological constant, extending beyond the parameters proposed by the no-hair theorem.The research provides insights into the swift decline of peaks linked to both large and small black holes, revealing new aspects of black hole transitional behaviors. In a three-dimensional space (for $d=3$) with a variable $k_1$ set to 1, and considering a $\Lambda$ term that adheres to SO(2) symmetry, the study uncovers a cusp catastrophe in the G-T function graph. This observation, within the specified metric, points to a distinct solution that characterizes the ``Phase Transition and Properties of Bose-Einstein Condensation" under specific conditions. Notably, this phase transition in Bose-Einstein condensation occurs due to the symmetry shift from SO(3) to SO(2).

 \text { KEYwords: } f(R) \text { gravity; thermodynamic geometry; black hole;Bose-Einstein Condensation. }

\end{abstract}

\maketitle

\section{Introduction}
Thermodynamic geometric analysis is an approach to exploring the thermodynamic attributes of black holes by examining their geometric traits and thermodynamic properties. Recently, researchers have begun to perform thermodynamic geometric analysis of 3D charged black holes under f(R) gravity.\cite{1,2,3,4,5,6}

3D-charged black holes represent a category of electrically charged black holes in a three-dimensional spacetime. Here, the electromagnetic repulsion from the charged constituents counteracts the gravitational attraction. The f(R) gravity theory tweaks Einstein's general relativity by implementing a fresh function of the Ricci scalar curvature. Such alterations are considered to furnish a fuller portrayal of gravity's behavior across both cosmological and quantum dimensions.

The thermodynamic attributes of black holes typically manifest in their entropy, temperature, and other thermodynamic variables. Under f(R) gravity, the thermodynamic features of 3D charged black holes are explored utilizing the geometric techniques developed by Ruppeiner and Quevedo. These methodologies consist of correlating the black hole's thermodynamic variables to a thermodynamic plane, symbolizing a geometric representation of the thermodynamic state space.

The Ruppeiner geometry approach pivots on the curvature concept, linking the curvature of the thermodynamic plane to the fluctuations in thermodynamic variables. In contrast, Quevedo's method operates on the information geometry principle, associating the thermodynamic plane metric with the Fisher information metric.\cite{4,5,6,7,8}

Applying these geometric methods to 3D-charged black holes in the realm of f(R) gravity has unveiled intriguing findings. For instance, it's been demonstrated that the Ruppeiner curvature scalar for these black holes is negative, pointing to repulsive interactions amid their microscopic components. This contrasts with the positive curvature scalar noted in non-spinning black holes under Einstein's general relativity.

Overall, studying 3D charged black holes in f(R) gravity via thermodynamic geometric analysis offers an innovative angle on black hole thermodynamics. This lens might clarify gravity's inherent nature and its varying behaviors, potentially offering insights into black hole physics and the broader universe structure.

The black hole thermodynamic phase transition, as viewed through thermodynamic geometry, serves as a pivotal method for phase transition research\cite{4,5,6,7,8}. Hence, we intend to design three distinct thermodynamic geometries for this specific black hole in the parameter space, rooted in the Hessian matrix: the Weinhold, Ruppeiner, and free energy geometries. For each geometry, their geometric standard curvatures will be computed separately. The scalar curvature's unique form will be thoroughly analyzed for divergence behavior, comparing the black hole's phase transition point and critical point to discern the link between the black hole's phase transition point and singular curvature behaviors.

The article then delves into two scenarios associated with the cosmological constant, with one intriguingly being non-constant. This non-constant variant is tied to f(R) gravity, the scalar field, or the electromagnetic tensor. The deemed cosmological constant is inherently linked to $\pi$. Liu\cite{8} further developed this, suggesting that the pressure term P might spawn fractals, potentially introducing extra dimensions. This interpretation isn't wholly in sync with f(R) gravity. In one scenario under f(R) gravity, geometric impacts emanate from the collective effect of all microscopic particles, making the cosmological constant genuinely constant. However, in another setting, it's connected to the scalar field.

In f(R) gravity's context, the 3D charged black hole solution is scrutinized. Emphasis is on the non-zero constant scalar curvature solution, probing the metric tensor adhering to the altered field equation. The black hole thermodynamics are assessed, examining their local and global stabilities without involving the cosmological constants. This analysis spans several f(R) models. The core disparities between the theories are underlined by juxtaposing our outcomes with general relativity. Our exploration unveils a profound thermodynamic phenomenology defining the f(R) gravity paradigm.

This article delves into 3D charged black holes within f(R) gravity, extracting the altered thermodynamic entropy. We investigate the thermodynamic quantities and the black holes' thermodynamic geometry. It's noted that the Hawking temperature inversely correlates with the event horizon radius. To assess the static black hole entropy changes due to thermal fluctuations in f(R) gravity, the Hawking temperature and unaltered specific heat expressions are employed. Our observations indicate that the G-T diagram suggests flatness when the cosmological constant omits negative power terms. However, a comet-shaped structure emerges in the G-T diagram when it includes these terms. Although defining these negative power terms might be daunting, this conclusion remains intact. Within the f(R) gravity framework, multiple 3D charged black holes appear to display traits akin to an ideal gas when the initial curvature scalar remains constant. Yet, when it's non-constant and the cosmological constant term has a negative power, these black holes might exhibit van der Waals gas-like attributes. We outline general solutions for non-negative power scenarios and specific solutions for negative power cases. Intriguingly, under certain conditions, one can observe a phase transition reminiscent of the Van der Waals gas for charged black holes under f(R) gravity. Beyond the event horizon, potential energy becomes independent of $r$ due to a barrier.
Despite this, the no-hair theorem links its components to $r$. Consequently, a van der Waals-like phase transition occurs, aligning the black hole's fate more with the cosmological constant than solely the no-hair theorem's parameters.

Bose-Einstein Condensate (BEC): This is a state of matter formed when many bosons (such as photons or atoms) cluster together at extremely low temperatures, exhibiting macroscopic quantum behaviors.SO(2) Symmetry: This describes a continuous symmetry involving rotations around a fixed axis. For instance, a disk rotating around its central axis maintains its physical properties due to its SO(2) symmetry.SO(3) Symmetry: This represents a continuous symmetry for all rotations in three-dimensional space. Rotations in any direction do not alter the physical laws; hence, the three-dimensional space possesses SO(3) symmetry. What is the connection between BEC and SO(2) and SO(3) symmetries?SO(2) Symmetry in BEC: When BEC is confined to a two-dimensional plane, it often exhibits SO(2) symmetry. This means that rotations in any direction on the plane do not change the physical properties of the BEC.\cite{6}

Evolution from SO(2) to SO(3) Symmetry: If a two-dimensional BEC is slowly released into three-dimensional space, its SO(2) symmetry gradually evolves into the more general SO(3) symmetry. This means that three-dimensional rotations are no longer confined to the plane, and the BEC remains invariant under rotations in any direction.

This evolutionary process involves complex quantum mechanical effects in condensed matter physics and requires advanced theoretical tools and experimental techniques for in-depth study. However, overall, it demonstrates how symmetry evolves with the changing dimensions and constraints of a physical system, presenting a fascinating physical phenomenon.

The article's layout is as follows: Section 2 delves into thermodynamic parameters in f(R) theory and entropy corrections. Section 3 explores the thermodynamics of 3D-charged black holes within an f(R) gravity backdrop. Section 4 concludes the phase Transition and Properties of Bose-Einstein Condensation in 3D Charged Black Holes. Finally, the paper concludes with a summary of our findings.

\section{Thermodynamic parameters and thermodynamic entropy correction in f(R) theory}

This section presents an overview of the formula for the thermodynamic entropy adjustment of black holes resulting from minor fluctuations around equilibrium. Initially, we'll set the stage by defining the state density at constant energy, using the natural units where $4 \pi=G=\hbar=c=1$ \cite{9,10}:
 \begin{equation}
\rho(E)=\frac{1}{2 \pi i} \int_{c-i \infty}^{c+i \infty} e^{\mathcal{S}(\beta)} d \beta.
\end{equation}

The precise entropy, denoted as $\mathcal{S}(\beta)=\log Z(\beta)+\beta E$, is influenced by the temperature $T(\beta^{-1})$ and isn't solely its equilibrium value. This entropy is the cumulative entropy of the thermodynamic system's smaller subsystems, which are sufficiently minor to be deemed at equilibrium. To delve into the nature of this exact entropy, we resolve the complex integral using the steepest descent method around the saddle point, $\beta_{0}=T_H^{-1}$, where $\left.\frac{\partial \mathcal{S}(\beta)}{\partial \beta}\right|_{\beta=\beta_{0}}=0$. A Taylor series expansion of the exact entropy around the point $\beta=\beta_{0}$ gives us:
\begin{equation}
\mathcal{S}(\beta)=\mathcal{S}_{0}+\frac{1}{2}\left(\beta-\beta_{0}\right)^{2}\left(\frac{\partial^{2} \mathcal{S}(\beta)}{\partial \beta^{2}}\right)_{\beta=\beta_{0}}+(\text { higher order terms })
\end{equation}
 and
 \begin{equation}
\rho(E)=\frac{e^{\mathcal{S}_{0}}}{2 \pi i} \int_{c-i \infty}^{c+i \infty} \exp \left[\frac{1}{2}\left(\beta-\beta_{0}\right)^{2}\left(\frac{\partial^{2} \mathcal{S}(\beta)}{\partial \beta^{2}}\right)_{\beta=\beta_{0}}\right] d \beta,
\end{equation}
we get that
\begin{equation}
\rho(E)=\frac{e^{\mathcal{S}_{0}}}{\sqrt{2 \pi\left(\frac{\partial^{2} \mathcal{S}(\beta)}{\partial \beta^{2}}\right)_{\beta=\beta_{0}}}},
\end{equation}
where $c=\beta_{0}$ and $\left.\frac{\partial^{2} \mathcal{S}(\beta)}{\partial \beta^{2}}\right|_{\beta=\beta_{0}}>0$ are chosen.

By leveraging the Wald relationship, which connects the Noether charge of the differential homeomorphism to the entropy of a generic spacetime with bifurcation surfaces, we propose a technique to extract the effective set of higher-order derivatives from the black hole entropy. Beginning with this entropy, we scrutinize the derivation procedure of the action functional \cite{11,12,13}. It's noteworthy that this paper asserts $4\pi$ to be 1.

When entropy $S$ is a transcendental number with respect to $\pi$, the expression of thermodynamic geometry metric is no longer applicable. If we extend it to the metric of $M$ with respect to $\Lambda_0$ and $r_{+}$, they are both functions of entropy. We then see the possibility of a Van der Waals gas phase transition.\cite{14,15}For specific details, see the appendix.

1. Metric:
\begin{equation}
\mathrm{d} M^2 = -e^{2 A_1} \mathrm{~d} \Lambda_0^2 + e^{-2 A_1} \mathrm{~d} r_{+}^2 + e^{2 A_2} \mathrm{~d} \theta^2 + e^{-2 A_2} \mathrm{~d} \varphi^2.
\end{equation} We set the cosmological constant as $\Lambda_0$.

2. Christoffel symbols:
\begin{equation}
\begin{aligned}
& \Gamma_{00}^1=A_1{ }^{\prime} e^{4 A_1}, \\
& \Gamma_{10}^0=A_1{ }^{\prime}, \\
& \Gamma_{11}^1=-A_1{ }^{\prime}, \\
& \Gamma_{21}^2=A_2{ }^{\prime}, \\
& \Gamma_{22}^1=-A_2{ }^{\prime} e^{2(A_2+A_1)}, \\
& \Gamma_{31}^3=-A_2{ }^{\prime}, \\
& \Gamma_{33}^1=A_2{ }^{\prime} e^{2(A_1-A_2)}.
\end{aligned}
\end{equation}

3. Ricci tensors:
\begin{equation}
\begin{aligned}
& R_{00}=(A_1^{\prime \prime} + A_1{ }^2 - A_1{ }^{\prime}(-A_1)^{\prime} + A_1{ }^{\prime} A_2{ }^{\prime} + A_1{ }^{\prime}(-A_2)^{\prime}) e^{4(A_1)}, \\
& R_{11}=A_1{ }^{\prime}(-A_1)^{\prime} + A_2{ }^{\prime}(-A_1)^{\prime} + -A_2{ }^{\prime}(-A_1)^{\prime} - A_2^{\prime \prime} - (-A_2)^{\prime \prime} - A_1{ }^2 - A_2{ }^2 - (-A_2)^{\prime 2} - A_1^{\prime \prime}, \\
& R_{22}=(-A_2^{\prime \prime} + A_2{ }^{\prime}(-A_1)^{\prime} - A_1{ }^{\prime}(A_2)^{\prime} - (-A_2)^{\prime} A_2{ }^{\prime} - A_2{ }^{\prime 2}) e^{2(A_1+A_2)}, \\
& R_{33}=(-(-A_2)^{\prime \prime} - A_2{ }^{\prime}(-A_1)^{\prime} - A_1{ }^{\prime}(-A_2)^{\prime} - A_2{ }^{\prime}(-A_2)^{\prime} - (-A_2)^{\prime 2}) e^{2(A_1-A_2)}.
\end{aligned}
\end{equation}

4. Ricci scalar:
\begin{equation}
\begin{array}{r}
R=(-2 A_1^{\prime \prime} - 2 A_1{ }^2 + 2 A_1{ }^{\prime}(-A_1)^{\prime} - 2 A_1{ }^{\prime} A_2{ }^{\prime} - 2 A_1{ }^{\prime}(-A_2)^{\prime} + 2 A_2{ }^{\prime}(-A_1)^{\prime} + 2 A_2{ }^{\prime}(-A_1)^{\prime} \\
\left. -2 A_2{ }^{\prime \prime} - 2(-A_2)^{\prime \prime} - 2 A_2{ }^{\prime 2} - 2(-A_2)^{\prime 2} - 2(-A_2)^{\prime} A_2{ }^{\prime}\right) e^{2 A_1}.
\end{array}
\end{equation}where $\mathrm{A}_1$ and $\mathrm{A}_2$ are modulo 1 or more.When the curvature of entropy $S$ with respect to $R$ geometry does not diverge, there is the possibility of divergence in the generalized solutions obtained.
We see the possibility of divergence in the curvature of thermodynamic $R$ geometry.

\section{Thermodynamic of RN black holes in $f(R)$ gravity background}
In this section, we briefly review the main features of the four-dimensional charged AdS black hole corresponding in the $f(R)$ gravity background with a constant Ricci scalar curvature\cite{13,14}. It is possible for the cosmological constant to be negative and the initial curvature scalar to be positive. In this case, the negative cosmological constant would tend to decelerate the universe's expansion, while the positive initial curvature scalar would tend to accelerate it. The overall effect would depend on the relative magnitudes of the two quantities.So in this article, we set the cosmological constant as
\begin{equation}
{\Lambda_0} = -{\Lambda}.
\end{equation}There, $\Lambda_0$ is represented as the cosmological constant term, and $\Lambda$ is represented as a constant that is positively correlated with the initial curvature scalar( ${\Lambda_0}/(4\pi) = -{\Lambda},$ when $4\pi=1$).

The action is given by
\begin{equation}
S=\int_{\mathcal{M}} d^{3} x \sqrt{-g}\left[f(R)-F_{\mu \nu} F^{\mu \nu}\right].
\end{equation}

\subsection{Black hole in the form of f(R) theory: $f(R)=R-\lambda \exp (-\xi R)+\kappa R^{n}+\eta \ln (R)$}
The metric form is(The solution of RN black hole has a non-constant or constant Ricci curvature) \cite{15,16,17,18,19,20,21,22,23,24,25,26,27,28,29,30}, where
\begin{equation}
\begin{aligned}
    \mathrm{d}s^{2} =-g(r) \mathrm{d} t^{2} 
    +{g(r)}^{- 1} \mathrm{~d} r^{2}
    +r^{2} d \theta^2,
\end{aligned}
\end{equation}
where $R_0(>0)$ is the cosmological constant.

\begin{equation}
g(r)=k_1-\frac{2\Lambda_0}{(d-1)(d-2)} r^{2}-\frac{M}{r^{d-3}}+\frac{ Q^{2}}{r^{d-2}}.
\end{equation}
\begin{equation}
\begin{aligned}
&\lambda=\frac{R+\kappa R^{n}-\left(R+n \kappa R^{n}\right) \ln R}{(1+\xi R \ln R) e^{ -\xi R}}, \\
&\eta=-\frac{(1+\xi R) R+(n+\xi R) \kappa R^{n}}{1+\xi R \ln R},
\end{aligned}
\end{equation}where $\xi$ is a free parameter, and ${k_1}$ is a constant; it can take the values 1, -1, or 0. $M$ represents the mass of the black hole, while $\Lambda$ and $Q$ represent the cosmological constant and the charge of the black hole, respectively. Our objective is to determine the entropy $S$.

To calculate the entropy of the black hole, we first need to compute its Hawking temperature. The Hawking temperature can be obtained by evaluating the effect of surface gravity, derived from evaluating the black hole metric at the event horizon. In this case, the event horizon is located at $r=r_+$.

By solving the equation $g\left(r_{+}\right)=0$, we can obtain the value of $r_+$ that satisfies the equation.

\subsubsection{When d=3, ${k_1}$=1}
The mathematical framework of this section is based on the SO(3) group. For more details, please refer to the appendix.

We have the following equations:\cite{31,32,33,34,35,36,37}

\begin{equation}
g(r)=-\Lambda_0 r^{2}-M +\frac{Q^{2}}{r}+1.
\end{equation}

At this point, ${r_+}$ is obtained :

\begin{equation}
\begin{aligned}
& \left\{\left\{r_+ = \frac{2^{1 / 3}(-3 \Lambda+3 M \Lambda)}{3 \Lambda\left(-27 Q^2 \Lambda^2+\sqrt{729 Q^4 \Lambda^4+4(-3 \Lambda+3 M \Lambda)^3}\right)^{1 / 3}}-\right.\right. \\
& \left.\frac{\left(-27 Q^2 \Lambda^2+\sqrt{729 Q^4 \Lambda^4+4(-3 \Lambda+3 M \Lambda)^3}\right)^{1 / 3}}{3 \times 2^{1 / 3} \Lambda}\right\}, \\
&
\end{aligned}
\end{equation}
where $r_+$ is the event horizon radius and the unique Killing horizon radius.

When g(r)=0,we get $\Lambda=\pi/{r^x}$,$x \in [0, 1]$
\begin{equation}
M=M_4=\Lambda r^{2} +\frac{Q^{2}}{r}+1.
\end{equation}

The pressure $P$ of a black hole and its volume $V$ are as follows:
\begin{equation}
P=\Lambda/(4\pi),V= 4\pi r^2.
\end{equation}

The calculation of the Hawking temperature using the conventional method is as follows:
\begin{equation}
T_4=\frac{\Lambda r_{+}}{2 \pi}-\frac{2Q^{2}}{ r_{+}^{2}}.
\end{equation}

The Gibbs free energy can be derived as:

\begin{equation}
G_4=-\Lambda r_{+}^{2}.
\end{equation}

We have obtained the partial derivative of $T_4$ with respect to $M$. This derivative is quite complex, but it is the crucial part we need to calculate $C_P$ and $C_V$.
Now, we can use the following formulas to compute $C_P$ and $C_V$ :
\begin{equation}
\begin{gathered}
C_P=\left(\frac{\partial M}{\partial T_4}\right)_P=\frac{1}{\left(\frac{\partial T_4}{\partial M}\right)_P} \\
C_V=\left(\frac{\partial M}{\partial T_4}\right)_V=\frac{1}{\left(\frac{\partial T_4}{\partial M}\right)_V},
\end{gathered}
\end{equation}
where, $\left(\frac{\partial T_4}{\partial M}\right)_P$ and $\left(\frac{\partial T_4}{\partial M}\right)_V$ are the partial derivatives we just computed.

\begin{equation}
C_P=C_V=\left(\frac{\partial T_4}{\partial M}\right)^{-1}
\end{equation}where 
\begin{equation}
\resizebox{\textwidth}{!}{$
\begin{aligned}
\frac{\partial T_4}{\partial M} = \left( \frac{\Lambda \pi}{2} + \frac{4 Q^2}{{r_+}^3} \right) \times \left( \frac{-2 2^{\frac{1}{3}} (-3 \Lambda + 3 \Lambda M)^3}{\sqrt{4 (-3 \Lambda + 3 \Lambda M)^3 + 729 \Lambda^4 Q^4} \left(-27 \Lambda^2 Q^2 + \sqrt{4 (-3 \Lambda + 3 \Lambda M)^3 + 729 \Lambda^4 Q^4}\right)^{\frac{4}{3}}} + \frac{2^{\frac{1}{3}}}{\left(-27 \Lambda^2 Q^2 + \sqrt{4 (-3 \Lambda + 3 \Lambda M)^3 + 729 \Lambda^4 Q^4}\right)^{\frac{1}{3}}} - \frac{2^{\frac{2}{3}} (-3 \Lambda + 3 \Lambda M)^2}{\sqrt{4 (-3 \Lambda + 3 \Lambda M)^3 + 729 \Lambda^4 Q^4} \left(-27 \Lambda^2 Q^2 + \sqrt{4 (-3 \Lambda + 3 \Lambda M)^3 + 729 \Lambda^4 Q^4}\right)^{\frac{2}{3}}} \right).
\end{aligned}$}
\end{equation}

The entropy for this BTZ-f(R) black hole solution is:

\begin{equation}
S=4 \pi r_{+}.
\end{equation}

\begin{figure}
\centering
\includegraphics[width=0.5\textwidth]{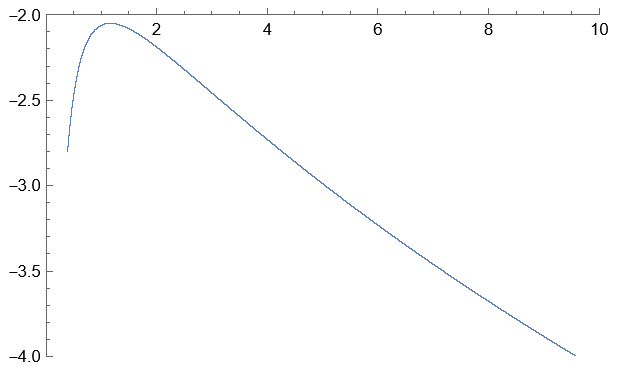}
\caption{At this time, Q=1,$\Lambda_0$=0.4. $G_4$ is the horizontal axis, and $T_4$ is the vertical axis.1. Curve Shape: The curve appears to have two distinct segments. The initial part (left) is more curved, while the latter part (right) is a straight line with a consistent negative slope.
2. X-Axis: It is labeled from an approximate range of 0 to 10.
3. Y-Axis: The visible values on the $y$-axis range from around -2.0 to -4.0 .
4. Inflection Point: There seems to be an inflection point or a point where the curve's nature changes, around the $x$-value of 2 . Before this point, the curve is declining at a decelerating rate, and after this point, the curve declines at a constant rate.
5. Overall Trend: The curve represents a decreasing trend throughout its length. The function seems to decrease more sharply after the $x$-value of 2 .
In summary, this plot represents a function that has a decreasing behavior across the displayed domain. Initially, the decrease is decelerating, but after a certain point (around $x=2$ ), it starts decreasing linearly. Further context is needed to provide insights about the data or the represented function.}
\label{10.png}
\end{figure}

\begin{figure}
\centering
\includegraphics[width=0.5\textwidth]{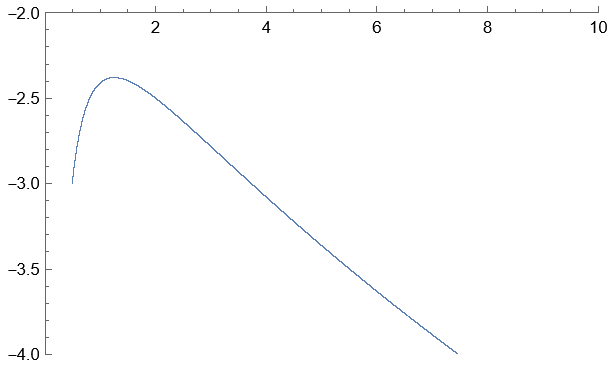}
\caption{At this time, Q=1,$\Lambda_0$=0.4. $G_4$ is the horizontal axis, and $T_4$ is the vertical axis.1. Curve Shape: The curve displays two distinct characteristics. The left portion exhibits a curved decline, transitioning to a linear descent as we progress to the right.
2. X-Axis: The $X$-axis is marked from an estimated range of 0 to 10.
3. Y-Axis: Visible values on the $Y$-axis extend from about -2.0 to -4.0 .
4. Transition Point: Around the $x$-value of 2 , the curve transitions from a curved decline to a straight-line descent. This suggests a change in the rate of decrease.
5. Overall Behavior: Throughout its span, the curve showcases a descending trend. The function experiences a more rapid decline post the $x$-value of 2 , suggesting a change in its behavior from a curved to a linear descent.
In essence, the plot illustrates a function that decreases across the shown domain. The decrease is curved initially but becomes linear after approximately $x=2$. Additional context or information would be required to provide deeper insights into the nature or implications of the depicted data.}
\label{11.png}
\end{figure}

\begin{figure}
\centering
\includegraphics[width=0.5\textwidth]{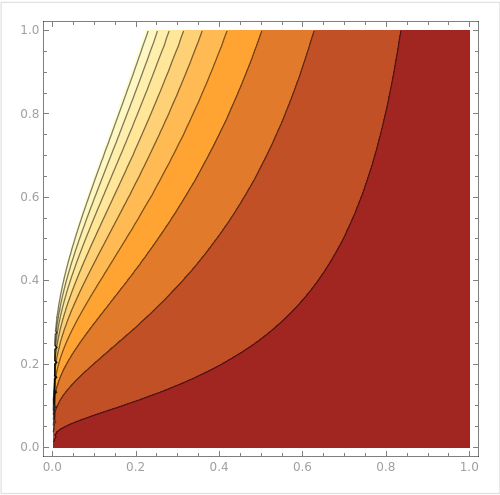}
\label{1.1111.jpg}
\caption{In this case, the horizontal axis is labeled as  `P', and the vertical axis is labeled as  `V'. This suggests that the graph could be representing a physical process or a state function where  `P' stands for pressure and  `V' stands for volume. This is reminiscent of diagrams used in thermodynamics, specifically those that depict properties of substances like gases where pressure and volume are related variables, such as in an isothermal or adiabatic process.}
\end{figure}

\begin{figure}
\centering
\includegraphics[width=0.5\textwidth]{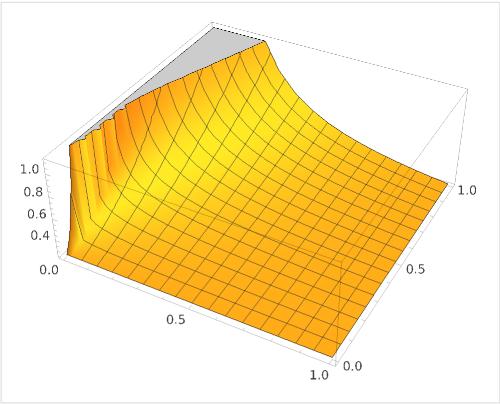}
\label{1.1112.jpg}
\caption{1.- $P$ is the vertical axis (usually the $z$-axis).
- $V$ is one of the horizontal axes (either the $x$-axis or the $y$-axis).
2. Since $x$ is a variable between 0 and 1 , the function might be exploring the influence of $x$ on the relationship between $P$ and $V$. As $x$ changes, the way the density $\Lambda$ changes with radius $r$ will also change, thus affecting the probability $P$.}
\end{figure}

We can express the metric as:

\begin{equation}
d s^{2}=-\frac{\partial^{2} {M_4}({r_+},\Lambda)}{\partial X^{\alpha} \partial X^{\beta}} \Delta X^{\alpha} \Delta X^{\beta},
\end{equation}

where
\begin{equation}
\frac{d^2{M_4}}{d {r_+}^2}=2\Lambda+\frac{2Q} {{r_+}^3},
\end{equation}
\begin{equation}
\frac{d^2{M_4}}{d {r_+}d {\Lambda}} = 2{r_+},
\end{equation}
\begin{equation}
\frac{d^2 {M_4}}{d\Lambda d\Lambda} = 0.
\end{equation}

The curvature scalar of the thermodynamic geometry can be calculated as:
\begin{equation}
R({S})= 0.
\end{equation}

In calculus, when we say differential, we are usually referring to a very small change, typically represented by symbols like $\Delta x$ or $d x$. These differentials have specific meanings in various computations, such as in derivatives and integrals.
$\pi$ is a specific constant, approximately 3.14159. It doesn't represent a change in itself, so it can't act as a differential on its own. However, in an expression or equation, $\pi$ can be combined with differentials. For instance, consider the change in the circumference of a circle with a change in its radius. If the radius changes by a small amount $d r$, then the change in circumference $d C$ can be expressed as:
\begin{equation}
d C=2 \pi d r.
\end{equation}
In this expression, $\pi$ is a constant, but $d r$ is a differential.
In summary, $\pi$ itself is not a differential, but it can be combined with differentials when expressing relationships of change.

If $\Lambda$ is a transcendental number with respect to $\pi$, then the scalar curvature of thermodynamic geometry is not necessarily zero.

\begin{figure}
\centering
\includegraphics[width=0.7\textwidth]{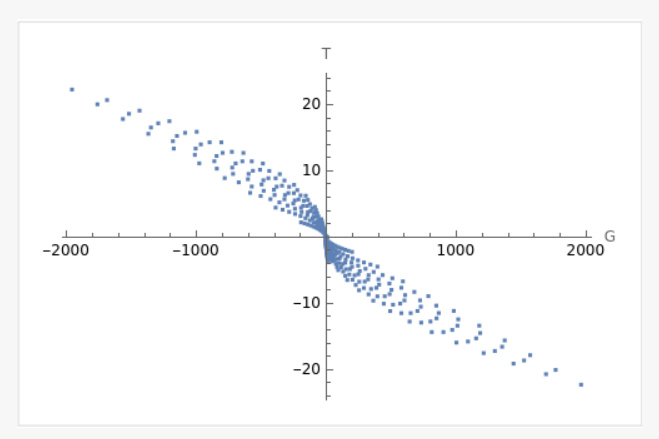}
\caption{At this time, Q=1,$\Lambda$ is transcendental number. $G_4$ is the horizontal axis, and $T_4$ is the vertical axis.}
\label{4.11.png}
\end{figure}
From the plot(section 3.1.1-FIG.5), it can be seen that the $G-T$ curve forms different shapes when $\Lambda$ and $r_{+}$ take different values. In some combinations of $\Lambda$ and $r_{+}$, the shape of the curve is similar to a swallowtail. These situations usually occur when the values of $\Lambda$ and $r_{+}$cause $G_4$ and $T_4$ to vary greatly within a certain range.

\begin{figure}
\centering
\includegraphics[width=0.7\textwidth]{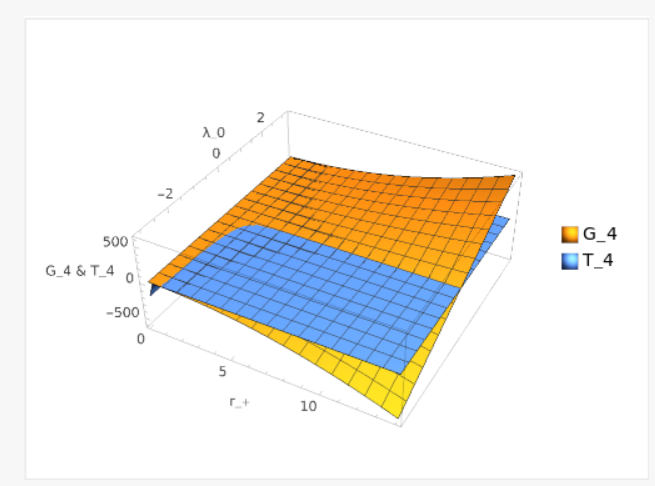}
\caption{At this time, Q=1,$\Lambda_0$ is $\pi$. }
\label{4.111.jpg}
\end{figure}
In the plot (Section 3.1.1-FIG.6), the $G-T$ curve exhibits varying shapes depending on the values of $\Lambda_0$ and $r_{+}$. For certain combinations of $\Lambda_0$ and $r_{+}$, the curve bears a resemblance to a swallowtail. Such configurations typically arise when the values of $\Lambda_0$ and $r_{+}$induce significant variations in $G_4$ and $T_4$ within a specific range.

We will now elucidate how the Schrödinger-like equation governs the radial behavior of the spatially confined non-minimally coupled mass scalar field configuration within the BTZ black hole spacetime. This is especially pertinent for the WKB analysis in the context of large masses. Specifically, employing the standard second-order WKB analysis of the radial equation, we derive the renowned discrete quantization condition. For this, we've utilized the integral relation:\cite{35,36,37}
\begin{equation}
\int_0^1 d x \sqrt{\frac{1}{x}-1}=\frac{\pi}{2} .
\end{equation}When $V(r \rightarrow+\infty)$ and $\mu^{\prime}=\frac{1}{n+\frac{1}{2}}$, the condition becomes:
\begin{equation}
\int_{\left(y^2\right)_{t-}}^{\left(y^2\right)_{t+}} d\left(y^2\right) \sqrt{\omega^2-V\left(y ; M, m, \mu^{\prime}, \alpha_1\right)}=\left(n+\frac{1}{2}\right) \frac{\pi \mu^{\prime}}{2}=\frac{\pi}{2} \quad ; \quad n=0,1,2, \ldots
\end{equation}Here, the integration boundaries $\left\{y_{t-}, y_{t+}\right\}$ in the WKB formula represent the classical turning points, where $V\left(y_{t-}\right)=V\left(y_{t+}\right)=0$. The resonant parameter $n$, which takes values in $\{0,1,2, \ldots\}$, defines the boundless discrete resonant spectrum $\left\{\alpha_n\left(\mu^{\prime}, m\right)\right\}_{n=0}^{n=\infty}$ of the black hole-field system.

By correlating the radial coordinates $y$ and $r$, the WKB resonance equation can be reformulated as:
\begin{equation}
\int_{r_t}^{r_{t+}} d r \frac{\sqrt{-V\left(r ; M, m, \mu^{\prime}, \alpha\right)}}{g(r)}=\left(n+\frac{1}{2}\right) \pi \quad ; \quad n=0,1,2, \ldots
\end{equation}which helps pinpoint the radial turning points $\left\{r_t, r_{t+1}\right\}$ of the combined black hole-field binding potential.

Given the existence of a potential barrier outside the event horizon, there's a point where the potential energy becomes independent of $r$. However, the no-hair theorem predicates that all three elements are dependent on $r$. In such scenarios, a van der Waals-like phase transition occurs, tying the fate of the black hole to the cosmological constant, rather than solely relying on the elements prescribed by the no-hair theorem.

Given the integral:
\begin{equation}
\int_0^1 d\left(\frac{r}{r_+}\right) \sqrt{\frac{r_+}{r}-1}=\left(n+\frac{1}{2}\right) \pi \quad ; \quad n=0,1,2, \ldots.
\end{equation}
First, we need to integrate the expression with respect to $r_+$. After obtaining the result, we can set it equal to $\frac{\Lambda}{2}$ and solve for the relationship between $r$ and $\Lambda=\pi$.

The relationship between $r$ and $\Lambda$ is given by:
\begin{equation}
\Lambda<4 r\left(\sqrt{-1+\frac{1}{r}}-\arctan \left(\sqrt{-1+\frac{1}{r}}\right)\right).
\end{equation}
This equation provides the value of $\Lambda$ in terms of $r$. If you need a specific inequality relationship between $r$ and $\Lambda$, please provide more details or constraints.

To determine whether $\Lambda$ is an increasing or decreasing function of $r$, we need to compute the derivative of $\Lambda$ with respect to $r$ and analyze its sign.
If $\frac{d \Lambda}{d r}>0$ for all $r$ in the domain of interest, then $\Lambda$ is an increasing function of $r$. Conversely, if $\frac{d \Lambda}{d r}<0$, then $\Lambda$ is a decreasing function of $r$.

\begin{equation}
\frac{d \Lambda}{d r}=\frac{-2\left(-1+r+2 \sqrt{-1+\frac{1}{r}} r \arctan \left(\sqrt{-1+\frac{1}{r}}\right)\right)}{\sqrt{-1+\frac{1}{r}} r}.
\end{equation}

For $r<1$ :
 The term $-1+r$ is negative.

The term $\sqrt{-1+\frac{1}{r}}$ is the square root of a positive number, so it's positive.

 The arctangent of a positive number is positive.

Given these observations, the sign of the numerator is determined by the combined effect of the terms.

The combined effect of the terms in the numerator is negative for $r<1$.
The denominator, $\sqrt{-1+\frac{1}{r}} r$, is positive for $r<1$ since the square root of a positive number is positive and $r$ is positive.
Given that the numerator is negative and the denominator is positive for $r<1$, the derivative $\frac{d \Lambda}{d r}$ is negative for $r<1$.
Therefore, $\Lambda$ is a decreasing function of $r$ for $r<1$.

We get that
\begin{equation}
\Lambda<4 {r_+}\left(\sqrt{-1+\frac{1}{r_+}}-\arctan \left(\sqrt{-1+\frac{1}{r_+}}\right)\right).
\end{equation}Considering the presence of a potential barrier beyond the event horizon, there emerges a juncture where the potential energy remains unaffected by $r$. Yet, the no-hair theorem asserts that all its three components are intrinsically linked to $r$. In these circumstances, a phase transition reminiscent of the van der Waals phenomenon takes place. This transition associates the destiny of the black hole more with the cosmological constant than just the parameters outlined by the no-hair theorem.

\subsubsection{When d=3, ${k_1}$=-1}
We have the following equations:

\begin{equation}
g(r)=-\Lambda_0 r^{2}-M +\frac{Q^{2}}{r}-1.
\end{equation}

When $g(r)=0$,we see that:

\begin{equation}
\begin{aligned}
& \left\{\left\{r_+ = \frac{2^{1 / 3}(3 \Lambda+3 M \Lambda)}{3 \Lambda\left(-27 Q^2 \Lambda^2+\sqrt{729 Q^4 \Lambda^4+4(3 \Lambda+3 M \Lambda)^3}\right)^{1 / 3}}-\right.\right. \\
& \left.\frac{\left(-27 Q^2 \Lambda^2+\sqrt{729 Q^4 \Lambda^4+4(3 \Lambda+3 M \Lambda)^3}\right)^{1 / 3}}{3 \times 2^{1 / 3} \Lambda}\right\}, \\
&
\end{aligned}
\end{equation}
where $r_+$ is the event horizon radius and the unique Killing horizon radius.

When $g(r)=0$, we get
\begin{equation}
M_5=\Lambda r^{2} +\frac{Q^{2}}{r}-1.
\end{equation}

The calculation of the Hawking temperature using the conventional method is as follows:
\begin{equation}
T_5=\frac{\Lambda r_{+}}{2 \pi}-\frac{2Q^{2}}{ r_{+}^{2}}.
\end{equation}

The Gibbs free energy can be derived as:

\begin{equation}
G_5=-\Lambda r_{+}^{2}.
\end{equation}

The entropy for this BTZ-f(R) black hole solution is:

\begin{equation}
S=4 \pi r_{+}.
\end{equation}

\begin{equation}
C_P=C_V=\left(\frac{\partial T_5}{\partial M}\right)^{-1}.
\end{equation}where 
\begin{equation}
\resizebox{\textwidth}{!}{$
\begin{aligned}
\frac{\partial T_5}{\partial M} = \left(\frac{\Lambda}{2 \pi} + 4 Q^2\right) \left( \frac{-432 2^{\frac{1}{3}} \Lambda^3}{\sqrt{864 \Lambda^3 + 729 \Lambda^4 Q^4} \left(-27 \Lambda^2 Q^2 + \sqrt{864 \Lambda^3 + 729 \Lambda^4 Q^4}\right)^{\frac{4}{3}}} + \frac{36 2^{\frac{2}{3}} \Lambda^2}{\sqrt{864 \Lambda^3 + 729 \Lambda^4 Q^4} \left(-27 \Lambda^2 Q^2 + \sqrt{864 \Lambda^3 + 729 \Lambda^4 Q^4}\right)^{\frac{2}{3}}} + \frac{2^{\frac{1}{3}}}{\left(-27 \Lambda^2 Q^2 + \sqrt{864 \Lambda^3 + 729 \Lambda^4 Q^4}\right)^{\frac{1}{3}}} \right).
\end{aligned}$}
\end{equation}

\begin{figure}
\centering
\includegraphics[width=0.5\textwidth]{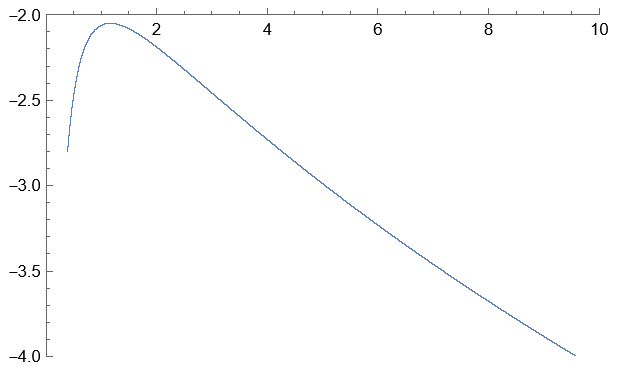}
\caption{At this time, Q=1, $\Lambda$=0.4. $G_5$ is the horizontal axis, and $T_5$ is the vertical axis.This image is consistent with the situation described in Section 3.1.1.}
\label{fig:10}
\end{figure}

\begin{figure}
\centering
\includegraphics[width=0.5\textwidth]{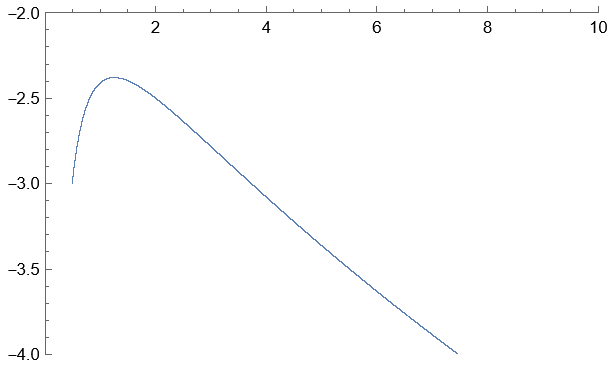}
\caption{At this time, Q=1, $\Lambda_0$=0.5. $G_5$ is the horizontal axis, and $T_5$ is the vertical axis.This image is consistent with the situation described in Section 3.1.1.}
\label{fig:11}
\end{figure}

The squared differential, $d s^2$, is given by:
\begin{equation}
d s^2=-\frac{\partial^2 M_5\left(r_{+}, \Lambda\right)}{\partial X^\alpha \partial X^\beta} \Delta X^\alpha \Delta X^\beta
\end{equation}
where:

1. The second derivative of $M_5$ with respect to $r_{+}$squared is:
\begin{equation}
\frac{d^2 M_5}{d r_{+}^2}=2 \Lambda+\frac{2 Q}{r_{+}^3}.
\end{equation}

2. The second derivative of $M_5$ with respect to $r_{+}$and $\Lambda$ is:
\begin{equation}
\frac{d^2 M_5}{d r_{+} d \Lambda}=2 r_{+}.
\end{equation}

3. The second derivative of $M_5$ with respect to $\Lambda$ squared is zero:
\begin{equation}
\frac{d^2 M_5}{d \Lambda^2}=0.
\end{equation}

Additionally, the curvature scalar of the thermodynamic geometry, represented as $R(S)$, evaluates to zero:
\begin{equation}
R(S)=0.
\end{equation}

In the realm of calculus, the term differential denotes a minute change. Symbols like $\Delta x$ or $d x$ typically represent this. These differentials play a significant role in various mathematical operations, such as derivatives and integrals. The constant, $\pi$,  is not representative of any change. Although $\pi$ can't act as a differential independently, it can be integrated into expressions with differentials. 

In conclusion, $\pi$ isn't a differential, but it can be merged with differentials to describe change relationships.

Furthermore, if $\Lambda$ is transcendental in relation to $\pi$, the scalar curvature of the thermodynamic geometry doesn't have to be zero.

\begin{figure}
\centering
\includegraphics[width=0.7\textwidth]{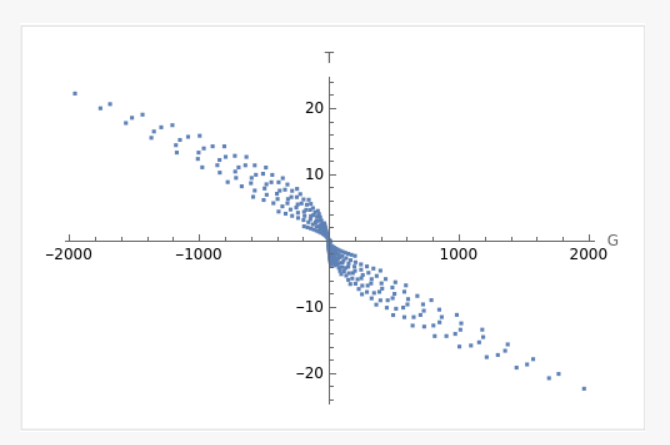}
\caption{At this time, Q=1,$\Lambda$ is transcendental number. $G_5$ is the horizontal axis, and $T_5$ is the vertical axis.}
\label{5.11.png}
\end{figure}From FIG.9, it can be seen that the $G-T$ graph will show different shapes, including a swallowtail shape, when $\Lambda$ and $r_{+}$take different values. This implies that phase transitions will occur in the system under certain values of $\Lambda$ and $r_{+}$, which is the physical meaning of the swallowtail shape.

\begin{figure}
\centering
\includegraphics[width=0.7\textwidth]{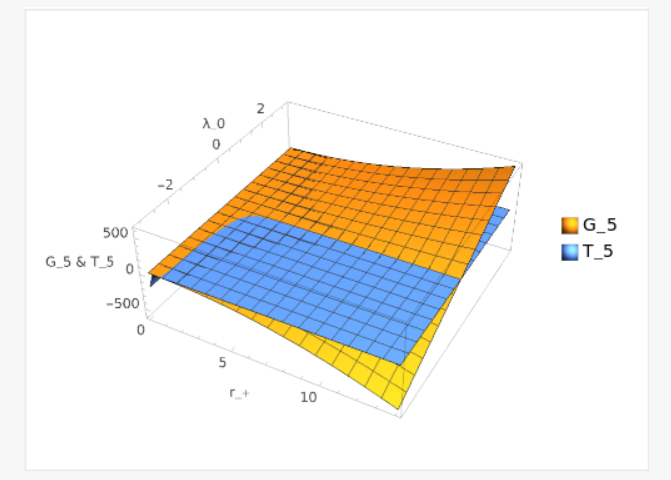}
\caption{At this time, Q=1,$\Lambda_0$ is $\pi$.}
\label{5.111.jpg}
\end{figure}In the plot (Section 3.1.2-(FIG.10)), the $G-T$ graph manifests various shapes, notably the distinctive swallowtail, depending on the values of $\Lambda_0$ and $r_{+}$. This suggests that phase transitions are likely to occur in the system for specific combinations of $\Lambda_0$ and $r_{+}$. The emergence of the swallowtail shape serves as a physical representation of these transitions.

We conclude the following, the conditions for generating van der Waals gas phase transitions are consistent with those in Section 3.1.1.It was found here that there is no van der Waals phase transition.

\subsubsection{When d=3, ${k_1}$=0}
We have the following equations:

\begin{equation}
g(r)=-\Lambda_0 r^{2}-M +\frac{Q^{2}}{r}.
\end{equation}

We can see:

\begin{equation}
\begin{aligned}
\left\{r_+= \frac{\sqrt[3]{\frac{2}{3}} M}{\sqrt[3]{\sqrt{3} \sqrt{4 \Lambda ^3 M^3+27 \Lambda ^4 Q^4}-9 \Lambda ^2 Q^2}}-\frac{\sqrt[3]{\sqrt{3} \sqrt{4 \Lambda ^3 M^3+27 \Lambda ^4 Q^4}-9 \Lambda ^2 Q^2}}{\sqrt[3]{2}* 3^{2/3} \Lambda }\right\},
\end{aligned}
\end{equation}
where $r_+$ is the event horizon radius and the unique Killing horizon radius.

When $g(r)=0$, we get

\begin{equation}
M=M_6=\Lambda r^{2} +\frac{Q^{2}}{r}.
\end{equation}

The calculation of the Hawking temperature using the conventional method is as follows:

\begin{equation}
T_6=\frac{\Lambda r_{+}}{2 \pi}-\frac{2Q^{2}}{ r_{+}^{2}}.
\end{equation}

The Gibbs free energy can be derived as:

\begin{equation}
G_6=-\Lambda r_{+}^{2}.
\end{equation}

The entropy for this BTZ-f(R) black hole solution is:

\begin{equation}
S=4 \pi r_{+}.
\end{equation}
\begin{equation}
C_P = C_V = \left( \frac{\partial T_6}{\partial M} \right)^{-1}
\end{equation}
\begin{equation}
\resizebox{\textwidth}{!}{$
\begin{aligned}
\frac{\partial T_6}{\partial M} = \left( \frac{\Lambda}{2 \pi} + \frac{4 Q^2}{r_+^3} \right) \times \left( \frac{-2 \times 2^{\frac{1}{3}} \times 3^{\frac{2}{3}} \Lambda^3 M^3}{\sqrt{4 \Lambda^3 M^3 + 27 \Lambda^4 Q^4} \left(-9 \Lambda^2 Q^2 + 3 \sqrt{4 \Lambda^3 M^3 + 27 \Lambda^4 Q^4}\right)^{\frac{4}{3}}} - \frac{2^{\frac{2}{3}} \times 3^{\frac{1}{3}} \Lambda^2 M^2}{\sqrt{4 \Lambda^3 M^3 + 27 \Lambda^4 Q^4} \left(-9 \Lambda^2 Q^2 + 3 \sqrt{4 \Lambda^3 M^3 + 27 \Lambda^4 Q^4}\right)^{\frac{2}{3}}} + \frac{2^{\frac{1}{3}}}{3^{\frac{1}{3}} \left(-9 \Lambda^2 Q^2 + 3 \sqrt{4 \Lambda^3 M^3 + 27 \Lambda^4 Q^4}\right)^{\frac{1}{3}}} \right)
\end{aligned}$}
\end{equation}

\begin{figure}
\centering
\includegraphics[width=0.5\textwidth]{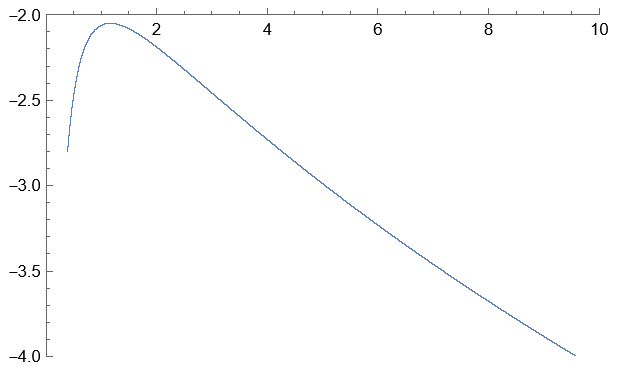}
\caption{At this time, Q=1, $\Lambda_0$=0.4. $G_6$ is the horizontal axis, and $T_6$ is the vertical axis.This image is consistent with the situation described in Section 3.1.1.}
\label{fig:10}
\end{figure}

\begin{figure}
\centering
\includegraphics[width=0.5\textwidth]{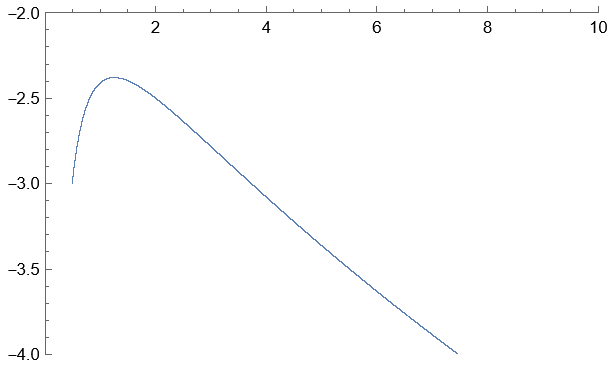}
\caption{At this time, Q=1, $\Lambda_0$=0.5. $G_6$ is the horizontal axis, and $T_6$ is the vertical axis.This image is consistent with the situation described in Section 3.1.1.}
\label{fig:11}
\end{figure}

The squared differential, denoted by \(d s^2\), is described by:
\begin{equation}
d s^2=-\frac{\partial^2 M_6\left(r_{+}, \Lambda\right)}{\partial X^\alpha \partial X^\beta} \Delta X^\alpha \Delta X^\beta
\end{equation}
with the following considerations:

\begin{enumerate}
    \item The second-order derivative of \(M_6\) concerning \(r_{+}\) squared is expressed as:
    \begin{equation}
    \frac{d^2 M_6}{d r_{+}^2}=2 \Lambda+\frac{2 Q}{r_{+}^3}.
    \end{equation}

    \item The mixed derivative of \(M_6\) with respect to both \(r_{+}\) and \(\Lambda\) is:
    \begin{equation}
    \frac{d^2 M_6}{d r_{+} d \Lambda}=2 r_{+}.
    \end{equation}

    \item The second-order derivative of \(M_6\) concerning \(\Lambda\) squared is null:
    \begin{equation}
    \frac{d^2 M_6}{d \Lambda^2}=0.
    \end{equation}
\end{enumerate}

Moreover, the curvature scalar of the thermodynamic geometry, denoted as \(R(S)\), is zero:
\begin{equation}
R(S)=0.
\end{equation}

In calculus, the term  `differential' refers to a small variation. Notations like \(\Delta x\) or \(d x\) are commonly used to represent such changes.  For instance, for a small alteration in a circle's radius, denoted by \(d r\), the variation in its circumference. Here, \(d r\) represents a differential, whereas \(\pi\) remains constant.

To sum up, while \(\pi\) isn't a differential, it can be combined with differentials to illustrate changes.

Additionally, if \(\Lambda\) is transcendental compared to \(\pi\), the scalar curvature of the thermodynamic geometry isn't necessarily zero.

\begin{figure}
\centering
\includegraphics[width=0.7\textwidth]{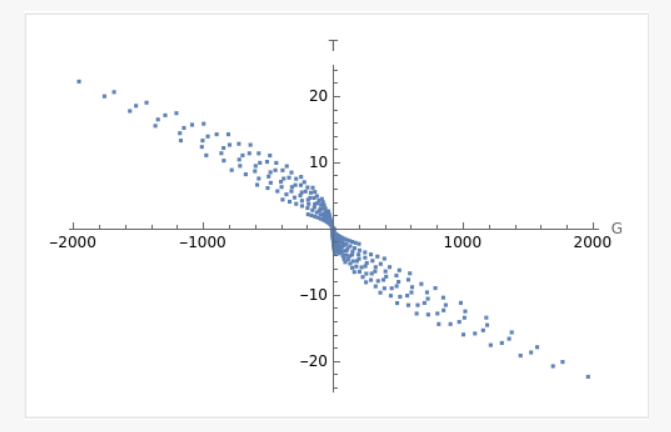}
\caption{At this time, Q=1. $G_6$ is the horizontal axis, and $T_6$ is the vertical axis.This is the plot of function $G$ versus $T_6$, where $G_6$ is the Gibbs free energy, $T_6$ is the temperature, $\Lambda_0$ is a transcendental number about $\pi$,and the range of $r_{+}$is from 1 to 14 .}
\label{6.11.jpg}
\end{figure}From the plot(Section 3.1.3-FIG.13), it can be seen that the $G-T$ graph will show different shapes, including a swallowtail shape, when $\Lambda$ and $r_{+}$take different values. This implies that phase transitions will occur in the system under certain values of $\Lambda$ and $r_{+}$, which is the physical meaning of the swallowtail shape.

\begin{figure}
\centering
\includegraphics[width=0.7\textwidth]{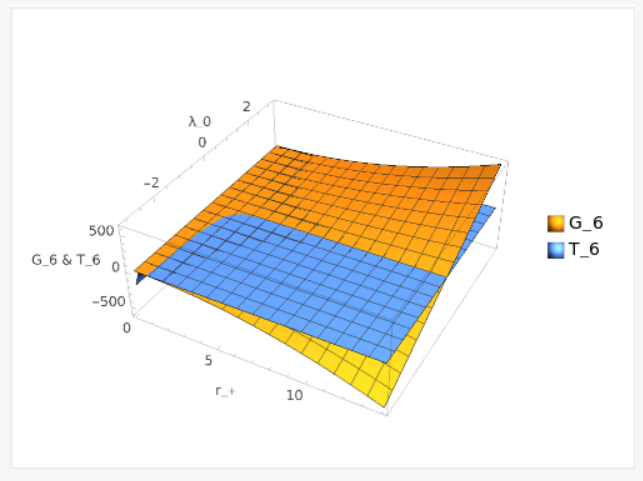}
\caption{The equations for $G_6$ and $T_6$ are the same as the ones provided for $G_6$ and $T_6$ as well as $G_5$ and $T_5$ in the previous queries. Given that $Q=1$ and $\Lambda_0=\pi$, the plot for $G_6$ and $T_6$ will be identical to the ones generated for the previous sets of equations.}
\label{6.111.jpg}
\end{figure}In the plot (Section 3.1.3-FIG.14), the $G-T$ graph exhibits various configurations, with the swallowtail shape being particularly prominent under certain combinations of $\Lambda$ and $r_{+}$. This suggests that the system undergoes phase transitions at specific values of $\Lambda$ and $r_{+}$. The presence of the swallowtail shape provides a physical interpretation of these transitions.

We have come to the following conclusion(from FIG.11,FIG.12,FIG.13,FIG.14): the conditions for generating van der Waals gas phase transitions align with those outlined in Section 3.1.1.It was found here that there is no van der Waals phase transition.

\subsection{Black hole in the form of f(R) theory: $$f(R)=-2 \eta M \ln (6 \Lambda_0+R)+R_{0}$$}
In the case where $\Phi(r)=0$, the charged $(2+1)$-dimensional solution under pure $f(R)$-gravity can be expressed using the metric given in follow,\cite{30,31,32,33,34}, where
\begin{equation}
g(r)=-\Lambda_0 r^{2}-M r-\frac{2 Q^{2}}{3 \eta r}.
\end{equation}

The two-dimensional line element is \cite{10,17}
\begin{equation}
d s^{2}=g(r) d \tau^{2}+\frac{d r^{2}}{g(r)}.
\end{equation}
To ensure consistency, we have made the assumption that both $\eta$ and $\xi$ do not take negative values.

At this point,we get that:
\begin{equation}
\begin{aligned}
&r_+=\frac{1}{3}\left(-\frac{M}{\Lambda_0}-\frac{M^{2} \eta}{\Lambda_0\left(M^{3} \eta^{3}+9 Q^{2} \eta^{2} \Lambda_0^{2}+3 \sqrt{2 M^{3} Q^{2} \eta^{5} \Lambda_0^{2}+9 Q^{4} \eta^{4} \Lambda_0^{4}}\right)^{1 / 3}}-\right.\\\
&\left.\frac{\left(M^{3} \eta^{3}+9 Q^{2} \eta^{2} \Lambda_0^{2}+3 \sqrt{2 M^{3} Q^{2} \eta^{5} \Lambda_0^{2}+9 Q^{4} \eta^{4} \Lambda_0^{4}}\right) ^{1 / 3}}{\eta \Lambda_0}\right),
\end{aligned}
\end{equation}where ${r_+}$ is the event horizon radius and the unique Killing horizon radius.

The calculation of the Hawking temperature using the conventional method is as follows:
\begin{equation}
T_7=-\frac{\Lambda_0 r_{+}}{2 \pi}-\frac{M}{4 \pi}+\frac{Q^{2}}{6 \pi \eta r_{+}^{2}}.
\end{equation}

Gibbs free energy can be derived as
\begin{equation}
G_7=\Lambda_0 r_{+}^{2}
\end{equation}
 The entropy for this BTZ-f(R) Black hole solution is
\begin{equation}
S=4 \pi r_{+}.
\end{equation}

\begin{equation}
{G_7}=H-{T_7}S={M_7}-{T_7} S=\frac{3 Q^2}{4 r_{+}}-\frac{2 P \pi r_{+}^3}{3}.
\end{equation}
\begin{figure}
  \centering
  \includegraphics[width=0.5\textwidth]{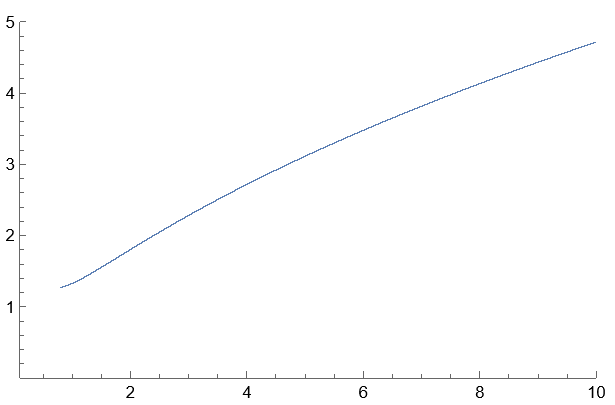}
  \caption{At this time, Q=1, P=0.8,M=1,$\eta=1$,$G_7$ is the horizontal axis, $T_7$ is the vertical axis}
  \label{8.png}
\end{figure}
\begin{figure}
  \centering
  \includegraphics[width=0.5\textwidth]{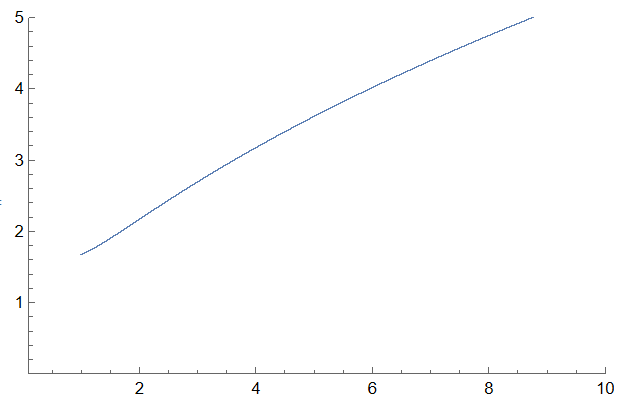}
  \caption{At this time, Q=1, P=1,M=1,$\eta=1$,$G_7$ is the horizontal axis, $T_7$ is the vertical axis}
  \label{9.png}
\end{figure}

Given:
\begin{equation}
M_7=\Lambda_0 r_{+}^2+\left(-\frac{\Lambda_0 r_{+}}{2 \pi}-\frac{M_7}{4 \pi}+\frac{Q^2}{6 \pi \eta r_{+}^2}\right)\left(4 \pi r_{+}\right).
\end{equation}

Expanding:
\begin{equation}
M_7=\Lambda_0 r_{+}^2-2 \Lambda_0 r_{+}^2-M_7 r_{+}+\frac{2 Q^2 r_{+}}{3 \eta r_{+}^2}.
\end{equation}

Combining like terms:
\begin{equation}
M_7=-\Lambda_0 r_{+}^2-M_7 r_{+}+\frac{2 Q^2 r_{+}}{3 \eta r_{+}^2}.
\end{equation}

Rearranging to isolate $M_7$ :
\begin{equation}
M_7\left(1+r_{+}\right)=-\Lambda_0 r_{+}^2+\frac{2 Q^2 r_{+}}{3 \eta r_{+}^2}.
\end{equation}

Finally:
\begin{equation}
M_7=\frac{-\Lambda_0 r_{+}^2+\frac{2 Q^2 r_{+}}{3 \eta r_{+}^2}}{1+r_{+}}.
\end{equation}
This is the expression for $M_7$ in terms of the given parameters.

We can express the metric as:
\begin{equation}
d s^{2}=-\frac{\partial^{2}  {M_7}(r_{+},\Lambda_0)}{\partial X^{\alpha} \partial X^{\beta}} \Delta X^{\alpha} \Delta X^{\beta},
\end{equation}where
\begin{equation}
R(S)=0.
\end{equation}The mentioned above do not involve any phase transition.

Comparing with Section 3.1.1.,we have reached the following conclusion: the conditions requisite for engendering van der Waals gas phase transitions are in alignment with those delineated in Section 3.1.1.It was found here that there is no van der Waals phase transition.

\section{Thermodynamics of 3D Charged $f(R)$ Black Holes}
In this scenario, the relevant equations \cite{31,32,33,34,35,36,37} are:
\begin{equation}
g(r)=-\Lambda_0 r^{2}-M +\frac{Q^{2}}{r}+1.
\end{equation}This is the only case of the metric on a three-dimensional charged black hole that satisfies the special phase transitions.

\subsection{The SO(3) Group}
The \textbf{SO(3)} group, known as the \textit{three-dimensional rotation group}, encapsulates all possible rotations within three-dimensional space. In the realm of physics, \textbf{SO(3)} is often linked with the quantum numbers of angular momentum, particularly in the context of describing fermions like electrons. For fermionic gases, the representations of the \textbf{SO(3)} group are intimately connected to the spin states of fermions, as reflected in their quantum statistics and the Pauli Exclusion Principle.\cite{38,39,40}

\subsection{The SO(2) Group and SO(3)/SO(2)}

\textbf{SO(2)}, the \textit{two-dimensional rotation group}, describes rotations in a two-dimensional plane. The subgroup \textbf{SO(3)/SO(2)} can be viewed as a subset of the \textbf{SO(3)} group, often employed in scenarios involving systems with reduced symmetry. In statistical mechanics, this reduction in symmetry might relate to the behavior of bosons, especially when considering phenomena like Bose-Einstein condensation.

These groups may manifest differently in the context of a 3D charged \textit{f(R)} black hole considered as a bosonic or fermionic gas. For a fermionic gas, the \textbf{SO(3)} group is closely tied to the spin of fermions and their quantum statistics. For a bosonic gas, \textbf{SO(3)/SO(2)} might be more pertinent, possibly relating to certain symmetry reductions or specific condensed states of the system.

\subsection{Symmetry Groups in Physics}

In physics, particularly within the frameworks of quantum and statistical mechanics, symmetry groups like \textbf{SO(3)} and \textbf{SO(2)} are often essential in describing system behaviors. These groups not only reflect the fundamental symmetrical properties of physical systems but also are closely associated with the quantum statistical characteristics of particles.

\begin{itemize}
    \item For \textit{fermionic gases}, the \textbf{SO(3)} group is typically related to the spin of fermions and the Pauli Exclusion Principle, crucial for describing the behavior of fermions such as electrons.
    \item For \textit{bosonic gases}, considerations of \textbf{SO(3)/SO(2)} may relate to phenomena like Bose-Einstein condensation, reflecting a reduction in system symmetry.
\end{itemize}

Overall, these groups play a pivotal role in describing the quantum statistical behaviors of bosonic and fermionic gases, especially in complex quantum systems like black holes. The choice of group depends on the specific physical properties of the system and the symmetries under consideration.

\subsection{The critical temperature of Bose-Einstein condensation of a Bose gas analogous to a black hole}
\subsubsection*{The critical temperature of Bose-Einstein condensation of a Bose gas}
For a 3D charged $f(R)$ black hole, the Gibbs free energy $G$ is given by:\cite{38,39}
\begin{equation}
G = -\Lambda r_{+}^{2}.
\end{equation}

If we consider a black hole molecule gas, the chemical potential $\mu$ is related to the Gibbs free energy $G$ by:
\begin{equation}
\mu = \left( \frac{\partial G}{\partial N} \right)_{S,V},
\end{equation}
where $N$ is the number of black hole molecules, $S$ is the entropy, and $V$ is the volume.

In many cases, the chemical potential can be considered as an approximation of the Gibbs function. This is because, for ideal gases or dilute solutions, the Gibbs function is directly proportional to the molar quantity of the substance. In such cases, the chemical potential is equal to the Gibbs function divided by the molar quantity of the substance.

For a Bose gas, the chemical potential is given by:
\begin{equation}
\mu = k_B T \ln\left( \frac{z}{e} \right),
\end{equation}
where $k_B$ is the Boltzmann constant, $T$ is the temperature, $z$ is the fugacity, and $e$ is the base of the natural logarithm.

Equating the two expressions for the chemical potential, we get:
\begin{equation}
k_B T \ln\left( \frac{z}{e} \right) = -\Lambda r_{+}^{2}.
\end{equation}

At the critical temperature $T_c$, the fugacity $z$ diverges, meaning the chemical potential becomes zero. Therefore, we have:
\begin{equation}
k_B T_c \ln\left( \frac{z}{e} \right) = -\Lambda r_{+c}^{2} = 0.
\end{equation}

Solving for $T_c$, we get:
\begin{equation}
T_c = \frac{\Lambda r_{+c}^{2}}{k_B \ln\left( \frac{z}{e} \right)}.
\end{equation}

 Just as particles in a BEC transition to a lower energy state en masse, the tunneling particles in our black hole model exhibit a collective behavior that can be analogized to this quantum phase transition. This analogy is particularly evident in the discussion of the phase and group velocities, where the interactions and emergent properties mirror those seen in BEC systems:

1.\textbf{Bose-Einstein Distribution}:

The average number of particles in a state with energy \( \epsilon \) is given by the Bose-Einstein distribution:
\begin{equation}
n(\epsilon) = \frac{1}{e^{(\epsilon - \mu)/kT} - 1},
\end{equation}
where \( \mu \) is the chemical potential, \( k \) is the Boltzmann constant, and \( T \) is the temperature.

2.\textbf{Particle Number Conservation}:

The total number of particles \( N \) in the system is the sum of particles in all energy states:
\begin{equation}
N = \sum_{\epsilon} n(\epsilon).
\end{equation}
In the continuous limit, the sum is replaced by an integral.

3.\textbf{Energy Density of Ideal Gas}:

For a three-dimensional ideal gas, the density of states \( g(\epsilon) \) as a function of energy is:
\begin{equation}
g(\epsilon) = \frac{V}{(2\pi)^2} \left( \frac{2m}{\hbar^2} \right)^{3/2} \epsilon^{1/2},
\end{equation}
where \( V \) is the volume, \( m \) is the particle mass, and \( \hbar \) is the reduced Planck constant.

4.\textbf{Solving for Condensation Temperature}:

Substituting \( g(\epsilon) \) into the particle conservation equation and setting the chemical potential \( \mu \rightarrow 0 \) (as in the case of BEC), we get:
\begin{equation}
N = \int_{0}^{\infty} \frac{g(\epsilon)}{e^{\epsilon/kT} - 1} d\epsilon.
\end{equation}
Solving this integral for temperature \( T \) at a given number of particles \( N \) and volume \( V \) yields the Bose-Einstein Condensation temperature.

5.\textbf{Approximate Expression for Temperature}:

In the context of a 3D charged black hole within an \( f(R) \) framework, we examine a solution characterized by an initial curvature \( R_0 \). The line element of this solution is expressed as(Satisfies the SO(3) group):\cite{30,31,32,33,34,35,36,37,38}
\begin{equation}
\begin{gathered}
\mathrm{d} s^{2}=-\left(1-\frac{2 M}{r}-\frac{R_{0} r^{2}}{12}+\frac{Q^2}{r^2}\right) \mathrm{d} t^{2}+ 
\left(1-\frac{2 M}{r}-\frac{R_{0} r^{2}}{12}+\frac{Q^2}{r^2}\right)^{-1} \mathrm{~d} r^{2}+r^{2} d \theta^2,
\end{gathered}
\end{equation}
where $R_{0}(>0)$ is the initial curvature.

To eliminate coordinate singularities in the measure, we introduce the generalized Painlev\'{e} coordinate transformation
\begin{equation}
dt_R = dt \mp \frac{\sqrt{1 -f}}{f}dr .
\end{equation}
The class Painlev\'{e} line elements of Reissner-Nordstr\"{o}m-de Sitter black hole-f(R) are obtained as follows
\begin{equation}
ds^2 = -f dt^2 \pm 2\sqrt{1 -f}dtdr +dr^2 +r^2(d\theta^2) .
\end{equation}

Now, let's use the new form of metric and obtain the radial geodesics of charged massive particles, which is different from the radial geodesics of uncharged massless particles that follow radial zero geodesics\cite{38}
\begin{equation}
\dot{r}=\frac{d r}{d t}=\pm 1 \mp \sqrt{1-f}.
\end{equation}
According to the de Broglie hypothesis, by the definition of the phase velocity $v_{p}$ and the group velocity $v_{g}$, we have
\begin{equation}
v_{p}=\frac{1}{2} v_{g}.
\end{equation}
Considering the instantaneous nature of the tunneling process and the Landau coordinate clock synchronization condition, the coordinate time difference for simultaneous events at different locations is:
\begin{equation}
d t=-\frac{g_{t r}}{g_{u}} d r_{c}, \quad(d \theta=0),
\end{equation}
where $d r_{c}$ is the position of the tunnel particle. So the group velocity can be expressed as(Hereinafter, we take the positive sign.)
\begin{equation}
v_{g}=\frac{d r_{c}}{d t}=-\frac{g_{t u}}{g_{t r}}=\pm \frac{r^{2}-2 M r+Q ^{2}-(R_0 / 12) r^{4}}{\sqrt{2 M r^{3}-Q^{2} r^{2}+(R_0 / 12) r^{6 }}}.
\end{equation}
So the phase velocity (radial geodesic) is
\begin{equation}
\dot{r}=v_{p}=-\frac{g_{ut}}{2 g_{t r}}=\pm \frac{r^{2}-2 M r+Q^{2}-( R_0 / 12) r^{4}}{2 \sqrt{2 M r^{3}-Q^{2} r^{2}+(R_0 / 12) r^{6}}},
\end{equation}
where the $+(-)$ sign represents the phase velocity of the charged particle passing through the EH (CH). In the process of the charged massive particle passing through the potential barrier, the self-interaction effect of the electromagnetic field on the emitted particle cannot be ignored, and the time component of the electromagnetic potential is
\begin{equation}
A_{l}=\pm \frac{Q}{r}.
\end{equation}
We will separately discuss Hawking radiation from the event and cosmic horizons and calculate the tunneling rate from each horizon. Given the complexity of metric tunneling radiation, we simplify our discussion to outgoing radiation, neglecting incoming radiation at the event horizon of a black hole.

\begin{equation}
\begin{aligned}
\operatorname{Im} S &=\operatorname{Im} \int_{r_{\mathrm{i}}}^{r_{\mathrm{f}}} \int_{M}^{M-\omega} \frac{1}{\dot{r}} \mathrm{~d}\left(M-\omega^{\prime}\right) \mathrm{d} r \\
&=\operatorname{Im} \int_{M}^{M-\omega} \int_{r_{\mathrm{i}}}^{r_{\mathrm{f}}} \frac{1}{1-\sqrt{1-f^{\prime}(r)}} \mathrm{d}\left(M-\omega^{\prime}\right) \mathrm{d} r,
\end{aligned}
\end{equation}

\begin{equation}
\frac{\partial}{\partial x^{j}}\left(-\frac{g_{0 i}}{g_{00}}\right)=\frac{\partial}{\partial x^{i}}\left(-\frac{g_{0 j}}{g_{00}}\right)
\end{equation}
\begin{equation}
\Gamma=\frac{\partial}{\partial x^{i}}\left(-\frac{g_{0 j}}{g_{00}}\right)=\frac{\partial}{\partial r}\frac{4 \sqrt{3} r^{2} \sqrt{\frac{-12 Q^{2}+24 M r+r^{4} R_0}{r^{2}}}}{12 Q^{2}-24 M r+12 r^{2}-r^{4}R_0}.
\end{equation}

\begin{equation}
\resizebox{\textwidth}{!}{$
\begin{aligned}
\Gamma = \frac{-4 \sqrt{3} r^2 (-24 M + 24 r - 4 r^3 R_0) \sqrt{\frac{-12 Q^2 + 24 M r + r^4 R_0}{r^2}}}{(12 Q^2 - 24 M r + 12 r^2 - r^4 R_0)^2} + \frac{8 \sqrt{3} r \sqrt{\frac{-12 Q^2 + 24 M r + r^4 R_0}{r^2}}}{12 Q^2 - 24 M r + 12 r^2 - r^4 R_0} + \frac{2 \sqrt{3} r^2 \left(\frac{24 M + 4 r^3 R_0}{r^2} - \frac{2 (-12 Q^2 + 24 M r + r^4 R_0)}{r^3}\right)}{(12 Q^2 - 24 M r + 12 r^2 - r^4 R_0) \sqrt{\frac{-12 Q^2 + 24 M r + r^4 R_0}{r^2}}}
\end{aligned}$}
\end{equation}

This is related to the emissivity of the tunneling particles
\begin{equation}
\Gamma \sim e^{-2 \operatorname{Im} S} .
\end{equation}
 $\Gamma$ gets,when ${R_0}=0$,
\begin{equation}
\resizebox{\textwidth}{!}{$
\begin{aligned}
\Gamma = \frac{-4 \sqrt{3} r^2 (-24 M + 24 r) \sqrt{\frac{-12 Q^2 + 24 M r}{r^2}}}{(12 Q^2 - 24 M r + 12 r^2)^2} + \frac{8 \sqrt{3} r \sqrt{\frac{-12 Q^2 + 24 M r}{r^2}}}{12 Q^2 - 24 M r + 12 r^2} + \frac{2 \sqrt{3} r^2 \left(\frac{24 M}{r^2} - \frac{2 (-12 Q^2 + 24 M r)}{r^3}\right)}{\sqrt{\frac{-12 Q^2 + 24 M r}{r^2}} (12 Q^2 - 24 M r + 12 r^2)}.
\end{aligned}$}
\end{equation}

When \( R_0 = 0 \) and \( r = 0 \), or simply when \( r = 0 \) alone, attempting to calculate \( \Gamma \) results in an indeterminate expression due to scenarios involving division by zero or negative powers of zero. Mathematically, this indicates that at the point where \( r = 0 \), \( \Gamma \) is undefined or singular. In the context of physics or geometry, this typically signifies that certain physical or geometrical properties are singular or discontinuous.

\subsection{Derivation of Bose-Einstein Condensation and Critical Temperature}
In this scenario, the relevant equations \cite{35,36,37} are(satisfying the $\mathrm{SO}(2)$ group form):
\begin{equation}
g(r)=-\Lambda_0 r^2-M+\frac{Q^2}{r}+1 .
\end{equation}

The Gibbs free energy of a black hole exhibits \(SO(3)\) group symmetry because it is a thermodynamic function under constant temperature and pressure conditions. Given that the conditions for Gibbs free energy are constant temperature and pressure, the cosmological constant under \(f(R)\) gravity satisfies \(SO(3)\) symmetry, aligning with the thin shell model.

The fundamental concept of the thin shell model involves considering a shell with thickness \(\delta\) and a distance \(\epsilon\) from the horizon, to study the entropy of the gas within the shell, and then taking the limits \(\delta \rightarrow 0\) and \(\epsilon \rightarrow 0\) to obtain the entropy of the horizon.
The Gibbs Free Energy G is\cite{37}
\begin{equation}
G = -\frac{2 \int_{0}^{\infty} \int_{\epsilon+{r_H}}^{\delta+\epsilon+{r_H}} \frac{\omega^2 \left( m^2 \left(-\Lambda_0 r^2 - M + \frac{Q^2}{r} + 1\right) \right)^{\frac{3}{2}}}{\left( e^{\beta \omega} - 1 \right) \left( -\Lambda_0 r^2 - M + \frac{Q^2}{r} + 1 \right)^2} \, dr \, d\omega}{3 \pi}
\end{equation}

\begin{equation}
G = -\frac{2 \int_{\epsilon+\mathrm{r_H}}^{\delta+\epsilon+\mathrm{r_H}} \frac{r^2 \left(\frac{m^2 \left(Q^2-r \left(\Lambda_0 r^2+M-1\right)\right)}{r}\right)^{3/4} \mathrm{If}\left[\Re(\beta)>0,\frac{2 \zeta (3)}{\beta^3},\int_{0}^{\infty} \frac{\omega^2}{e^{\beta \omega}-1}, \mathrm{d}\omega, \mathrm{Assumptions} \to \Re(\beta)\leq 0\right]}{\left(Q^2-r \left(\Lambda_0 r^2+M-1\right)\right)^2} \, \mathrm{d}r}{3 \pi }
\end{equation}
Simplification yields G is negative.
\begin{align*}
r_H =  \text{Horizon radius} \\
\epsilon =  \text{Offset from } r_H \\
\delta =  \text{Radial range of the integral}
\end{align*}
The Gibbs free energy, denoted as $G$, is defined by the equation:
\begin{equation}
G = H - TS, 
\end{equation}
where $H$ represents the enthalpy of the system, $T$ is the absolute temperature, and $S$ is the entropy of the system.

This concept is applicable to chemical reactions and phase transitions occurring under normal temperature and pressure conditions, specifically under constant pressure and temperature.

A process is considered spontaneous if the Gibbs free energy of the system decreases:
\begin{equation}
\Delta G < 0. 
\end{equation}
The main focus is on the energy changes in systems under normal pressure.

The formulas for the black hole Gibbs function are based on the thin shell model. By considering a thin shell at a specific distance outside the event horizon, these formulas aim to study the entropy of the gas within the shell. The entropy of the event horizon is obtained by allowing the thickness and distance of the shell to approach zero. This methodology is commonly used in theoretical physics, especially in exploring the thermodynamic properties of black holes.

The provided formulas for Gibbs free energy ($G$) differ in their expression form and the integral variables involved. They incorporate various physical quantities such as angular frequency ($\omega$), mass ($m$), the cosmological constant ($\Lambda_0$), black hole mass ($M$), charge ($Q$), and inverse temperature ($\beta$), constructed based on specific physical scenarios and assumptions.

We will now elucidate how the Schrödinger-like equation governs the radial behavior of the spatially confined, nonminimally coupled, mass scalar field configuration within the 3D charged black hole spacetime. This is especially pertinent for the WKB analysis in the context of large masses. Specifically, employing the standard second-order WKB analysis of the radial equation, we derive the renowned discrete quantization condition. For this, we have utilized the integral relation:\cite{35,36,37}
\begin{equation}
\int_0^1 d x \sqrt{\frac{1}{x}-1}=\frac{\pi}{2} .
\end{equation}
When $V(r \rightarrow+\infty)$ and $\mu^{\prime}=\frac{1}{n+\frac{1}{2}}$, the condition becomes:
\begin{equation}
\int_{\left(y^2\right)_{t-}}^{\left(y^2\right)_{t+}} d\left(y^2\right) \sqrt{\omega^2-V\left(y ; M, m, \mu^{\prime}, \alpha_1\right)}=\left(n+\frac{1}{2}\right) \frac{\pi \mu^{\prime}}{2}=\frac{\pi}{2} \quad ; \quad n=0,1,2, \ldots.
\end{equation}

Here, the integration boundaries $\left\{y_{t-}, y_{t+}\right\}$ in the WKB formula represent the classical turning points, where $V\left(y_{t-}\right)=$ $V\left(y_{t+}\right)=0$. The resonant parameter $n$, which takes values in $\{0,1,2, \ldots\}$, defines the boundless discrete resonant spectrum $\left\{\alpha_n\left(\mu^{\prime}, m\right)\right\}_{n=0}^{n=\infty}$ of the black hole-field system.
By correlating the radial coordinates $y$ and $r$, the WKB resonance equation can be reformulated as:
\begin{equation}
\int_{r_{t-}}^{r_{t+}} d r \frac{\sqrt{-V\left(r ; M, m, \mu^{\prime}, \alpha\right)}}{g(r)}=\left(n+\frac{1}{2}\right) \pi \quad ; \quad n=0,1,2, \ldots.
\end{equation}
which helps pinpoint the radial turning points $\left\{r_{t-}, r_{t+}\right\}$ of the combined black hole-field binding potential.

1. \textbf{Derivation of the Energy-Momentum Relation}

Within the framework of quantum field theory in curved spacetime, the dynamics of bosons are described by the Klein-Gordon equation. Considering the case of massless particles and using the WKB (Wentzel-Kramers-Brillouin) approximation, we obtain the energy-momentum relation for bosons as:\cite{37}
\begin{equation}
    p = \frac{(1 \pm \sqrt{1-f}) \hbar \omega}{f},
\end{equation}
where $f$ represents a factor related to gravity, $\hbar$ is the reduced Planck's constant, and $\omega$ is the angular frequency. In the special case where the shell approaches the horizon ($f \rightarrow 0$), the energy-momentum relation simplifies to:
\begin{equation}
    p = \frac{2 \hbar \omega}{f}.
\end{equation}

2.\textbf{ Calculation of the Phase Space Degeneracy}

To delve deeper into the thermodynamic properties of a boson gas in the shell, a 6-dimensional phase space description is introduced. Applying the generalized uncertainty principle to the smallest phase cell, we derive the expression for phase space degeneracy as:
\begin{equation}
    \omega_l = \frac{\mathrm{d} x \mathrm{~d} y \mathrm{~d} z \mathrm{~d} p_x \mathrm{~d} p_y \mathrm{~d} p_z}{\hbar^3\left(1+\frac{\alpha}{\hbar^2} p^2\right)^3},
\end{equation}
where $\alpha$ is a physical quantity characterizing the geometric properties of the phase space. In spherical coordinates, this degeneracy expression can be rewritten as:
\begin{equation}
    \omega_l = \frac{(4 \pi)^2 r^2 p^2 \mathrm{~d} r \mathrm{~d} p}{\hbar^3\left(1+\frac{\alpha}{\hbar^2} p^2\right)^3}.
\end{equation}

3. \textbf{Definition of Modified Thermodynamic Quantities}

Considering a thermodynamic system in equilibrium composed of a black hole and an external shell, we define modified thermodynamic quantities using the Bose statistical method. The total energy and number of particles are given by the following integrals:
\begin{equation}
    \mathcal{E} = \int_0^{\infty} \omega_l E(p) \mathrm{d} p,
\end{equation}
\begin{equation}
    \mathcal{N} = \int_0^{\infty} \omega_l \mathrm{d} p,
\end{equation}
where $E(p)$ represents the energy of a boson.

4.\textbf{ Form of the Modified Metric}

To account for the thermodynamic properties of the system, we define a modified metric form to describe the spacetime structure around the black hole:
\begin{equation}
    \mathrm{d}s^2 = -g(r) \mathrm{d} t^{2} + {g(r)}^{-1} \mathrm{d} r^{2} + r^{2} d\theta^2,
\end{equation}
where $g(r)$ is the metric function determined by the modified thermodynamic quantities and equations.

5. \textbf{Description of Bose-Einstein Condensation}

At sufficiently low temperatures, bosons undergo condensation, i.e., a large number of bosons gather in the same quantum state. Within the framework of modified thermodynamics and metric, the Bose-Einstein condensation can be described as:
\begin{equation}
    \mathcal{N}_0 = \int_0^{p_c} \omega_l \mathrm{d} p,
\end{equation}
where $p_c$ is the critical momentum for condensation. By calculating the modified energy spectrum $E(p) = \frac{\sqrt{p^2+m^2}}{\sqrt{-g(r)}}$ and substituting it into the Bose-Einstein condensation equation, we obtain the expression for the critical temperature of condensation:
\begin{equation}
    k_B T_c = \frac{2 \hbar^2}{m} \left( \frac{\mathcal{N}_0}{\zeta(3)} \right)^{1/3},
\end{equation}
where $k_B$ is the Boltzmann constant, $m$ is the mass of the boson, and $\zeta(3)$ is a special value of the Riemann zeta function. This expression reveals the relationship between the critical temperature of Bose-Einstein condensation and the system parameters, providing a theoretical foundation for understanding and studying the condensation phenomenon.

In crafting the Bose-Einstein Condensate (BEC) Wave Function Equation, we observe the phase transition characteristic of Bose-Einstein condensation emerging through the intricate disruption of SO(3)/SO(2) symmetry.

The Gross-Pitaevskii equation is commonly used to describe the BEC wave function. It's a nonlinear Schrödinger equation given by:
\begin{equation}
\begin{aligned}
i \hbar \frac{\partial \psi}{\partial t} &= \left( -\frac{\hbar^2}{2 m} \nabla^2 + V + g |\psi|^2 \right) \psi, \\
V &= -\Lambda_1 r^2 - M + \frac{Q^2}{r} + 1 ,\\
T_c &= \frac{2 \pi \hbar^2 n_c}{m k_B} ,\\
\mu &= -\Lambda_1 r^2 - M + \frac{Q^2}{r} + 1 + g |\psi|^2 ,\\
\mu \psi&=-\frac{\hbar^2}{2 m} \nabla^2 \psi + V \psi ,\\
\mu_c &= -\Lambda_1 r_c^2 - M + \frac{Q^2}{r_c} + 1 + g |\psi_c|^2, \\
T_c &= \frac{2 \pi \hbar^2 \mu_c}{m k_B}
\end{aligned}
\end{equation}

- $\Lambda_1$ : Cosmological constant related to the black hole spacetime.$\Lambda_1$ satisfies the SO(2) group form of $\Lambda_0$ in Section 2.3.1.

- $M$ : Mass parameter of the black hole.

- $Q$ : Charge of the black hole.

- $r_c$ : Critical radius at which the phase transition occurs.

- $g$ : Interaction strength in the BEC.

- $\mu$ :  The chemical potential and can also be regarded as the Gibbs function.

- $\left|\psi_c\right|^2$ : Density of the BEC at the critical point.

-$T_c$ : The critical temperature in the BEC.

\textbf{Algebraic Structures of Hawking Radiation Divergence Rate $\Gamma$ and Bose-Einstein Condensation Equation Degree of Freedom $g$}

The divergence rate $\Gamma$ of Hawking radiation can be represented as:
\begin{equation}
\Gamma = \frac{k^2 T}{2 \pi \hbar} \sum_{l=0}^{\infty} (2l + 1) e^{-2 \pi l / kT}
\end{equation}
where:
\begin{itemize}
    \item $k$ is the Boltzmann constant
    \item $T$ is the temperature of the black hole
    \item $\hbar$ is the reduced Planck constant
    \item $l$ is the quantum number of the black hole
\end{itemize}
\begin{equation}
\Gamma = \frac{k^2 T}{2 \pi \hbar} \left( \frac{1}{1 - x} \right),x = e^{-2 \pi / kT},\Gamma = \frac{k^2 T}{2 \pi \hbar} \sum_{n=0}^{\infty} x^n
\end{equation}

The Bose-Einstein condensation equation is:
\begin{equation}
\frac{\partial \rho}{\partial t} = - \frac{\nabla^2 \rho}{2m} + V(\mathbf{r}) \rho - g \rho^2
\end{equation}
where:
\begin{itemize}
    \item $\rho$ is the density of the Bose-Einstein condensate
    \item $t$ is time
    \item $m$ is the mass of the particle
    \item $V(\mathbf{r})$ is the external potential energy
    \item $g$ is the two-body interaction parameter
\end{itemize}

\begin{equation}
\frac{\partial \rho}{\partial t} = - \frac{\nabla^2 \rho}{2m} + \left( V(\mathbf{r}) - g \rho \right) \rho.
\end{equation}
\begin{equation}
\mu = V(\mathbf{r}) - g \rho,\rho = \frac{1}{e^{\beta (\mu - E)} + 1},\rho = \frac{1}{e^{\beta \mu} + 1},x = e^{-\beta \mu},\rho = \frac{1}{x + 1},\rho = \frac{1}{1 + e^{\beta \mu}}.
\end{equation}

By comparing the algebraic structures of the Hawking radiation divergence rate $\Gamma$ and the degree of freedom $g$ of the Bose-Einstein condensation equation, we find the following isomorphism:
\begin{equation}
\Gamma \sim \frac{1}{1 - e^{-2 \pi / kT}} \sim \frac{1}{x}
\end{equation}
\begin{equation}
g \sim \frac{1}{\beta \mu} \cdot \ln \left( \frac{\rho}{1 - \rho} \right) \sim \ln \left( \frac{1}{x} \right)
\end{equation}

This isomorphism indicates that the Hawking radiation divergence rate $\Gamma$ and the degree of freedom $g$ of the Bose-Einstein condensation equation have the same algebraic form. This means they can be viewed as two different aspects of the same mathematical object.

Under the background of a black hole, the emissivity of tunneling particles  $\Gamma$ and the interaction strength $g$ in the BEC can indeed be equal. This is because the gravitational field of the black hole alters the properties of the BEC, making it more susceptible to tunneling phenomena. 

Emissivity is the emission rate of the tunneling particles, Interaction strength is the interaction strength of the BEC, G is the gravitational constant, M is the mass of the black hole, and r is the distance between the particles and the black hole.

From the formula, it is evident that emissivity can be directly proportional to interaction strength. When the interaction strength is sufficiently large, the emissivity can reach 1, meaning that all particles have the potential to tunnel through.

One example to consider is a BEC composed of superfluids like helium. The interaction strength of superfluid helium is very high, so under the background of a black hole, the emissivity can reach 1. This implies that near a black hole, superfluid helium experiences strong tunneling, leading to a large number of particles escaping the black hole.

However, it is crucial to note that emissivity being equal to interaction strength is only a limiting case. In practical scenarios, factors like the temperature of the BEC and the rotational speed of the black hole limit emissivity.

At this juncture, we embrace the Gaussian unit system, where we set $4\pi = 1$, and introduce a specific scenario that meets the criteria of the SO(2) group.
\begin{equation}
\begin{aligned}
\Lambda_0(r) := \frac{1}{r^3} - \frac{5}{r^2} + 5,\\
G(r_+) := \Lambda_0(r_+) r_+^2,\\
T(r_+) := -\Lambda_0(r_+) {2r_+} - \frac{1}{r_+^2}.
\end{aligned}
\end{equation}
\begin{figure}
\centering
\includegraphics[width=0.7\textwidth]{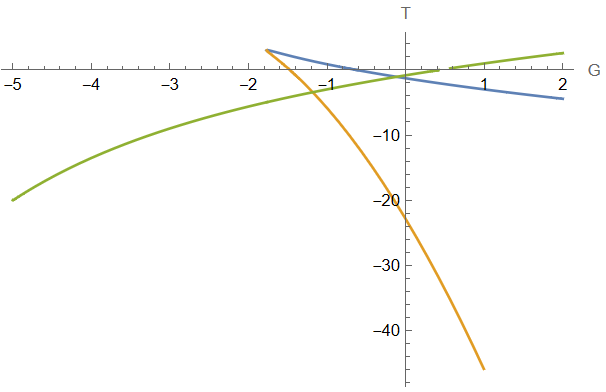}
\caption{At this time, Q=1,$\Lambda_0(r_+)= \frac{1}{r_+^3} - \frac{5}{r_+^2} + 5$. It resembled a Bose gas more closely at that time, undergoing Bose gas phase transitions (such as Bose-Einstein condensation).}
\label{4.1.jpg}
\end{figure}

When the $\Lambda$ term satisfies SO(2) group symmetry(FIG.5), and under this metric, the presence of a cusp catastrophe in the G-T function graph is observed, it can be demonstrated that under specific conditions, there exists a particular solution indicative of a Bose gas phase transition.When Q=0, we see that the black hole phase transition is analogous to the case of the Bose gas phase transition.

\subsubsection*{Advanced Analysis of Quantum Statistical Mechanics of Black Hole Quantum Statistical gas}
The grand canonical partition function $\mathcal{Z}$ for the black hole molecule gas is formulated as:
\begin{equation}
\mathcal{Z} = \mathrm{Tr} \left( e^{-\beta(\hat{H} - \mu \hat{N})} \right),
\end{equation}
where $\hat{H}$ is the Hamiltonian operator, $\hat{N}$ is the number operator, $\beta = 1/k_B T$, and $\mu$ is the chemical potential.

The effective action $S_{\text{eff}}$ for the black hole molecule gas within a quantum field theory framework is given by:
\begin{equation}
S_{\text{eff}}[\psi, \psi^*] = \int d^4x \left( \frac{\hbar^2}{2m}|\nabla \psi|^2 + V_{\text{eff}}(r, |\psi|^2) - \mu |\psi|^2 \right),
\end{equation}
with $\psi$ representing the complex field associated with the black hole molecules and $V_{\text{eff}}$ the effective potential.

The partition function in the path integral formalism is:
\begin{equation}
\mathcal{Z} = \int \mathcal{D}[\psi] \mathcal{D}[\psi^*] e^{-S_{\text{eff}}[\psi, \psi^*]/\hbar},
\end{equation}
highlighting the role of quantum fluctuations in the phase transition.

A renormalization group (RG) analysis elucidates the phase transition's scaling behaviors and universality classes.

\subsection{The Fermi energy level and Fermi degeneracy pressure of a Fermi gas analogous to a black hole}
The Fermi-Dirac distribution function:
\begin{equation}
f(E) = \frac{1}{e^{\frac{E - \mu}{kT}} + 1}.
\end{equation}
In this equation:

$f(E)$ represents the probability of a state with energy E being occupied.

E is the energy of the particle.

$\mu$ is the chemical potential known as the Fermi level.

$k$ is the Boltzmann constant.

T is the absolute temperature.

This distribution is crucial in quantum mechanics for describing the statistical behavior of fermions, such as electrons, especially considering the Pauli exclusion principle, which states that no two fermions can occupy the same quantum state simultaneously.

For a free electron gas, the Fermi energy can be calculated using the formula:\cite{38,39}
\begin{equation}
E_F = \frac{\hbar^2}{2m}(3\pi^2n)^{2/3}.
\end{equation}
where $h$ is the Planck constant and $m$ is the mass of the fermion.

Fermi degeneracy pressure is the pressure exerted by a Fermi gas due to quantum effects. For an electron gas at absolute zero, it can be expressed as:
\begin{equation}
P_F = \frac{(3\pi^2)^{2/3}\hbar^2}{5m}(n^{5/3}).
\end{equation}
The symbols have the same meaning as in the formula for Fermi energy.

These formulas apply to an ideal Fermi gas model and are particularly important in low-temperature conditions, such as in solid-state physics and stellar physics. Real-world systems might deviate due to electron interactions, lattice structures, and other factors.

The rotational invariance of the SO(3) group plays a crucial role in understanding the fundamental properties of Fermi gases. This aspect is particularly significant as it pertains to the symmetry of quantum states, a core concept in quantum mechanics. For Fermi gases, especially for spin-1/2 particles like electrons, their wave functions adhere to the rotational invariance of the SO(3) group, affecting their distribution in quantum states.

The calculation of Fermi energy levels and Fermi degeneracy pressure relies on Fermi-Dirac statistics, a form of quantum statistical distribution. This distribution takes into account the Pauli Exclusion Principle, which states that no two fermions can occupy the same quantum state. The rotational invariance of the SO(3) group influences the form of fermions' (such as electrons) wave functions, consequently affecting their distribution in energy states.

The rotational invariance of the SO(3) group is a key factor in understanding the behavior of Fermi gases. It is closely related to the calculation of Fermi energy levels and Fermi degeneracy pressure. However, it is not the sole necessary condition, as the calculation of these aspects also depends on other concepts in quantum statistics, such as the Pauli Exclusion Principle and Fermi-Dirac statistics.

The SO(2) group can be represented as the set of all 2D rotation matrices that preserve the origin. A typical SO(2) matrix is given by:\cite{38,39,40}
\begin{equation}
R(\theta) = \begin{pmatrix}
\cos\theta & -\sin\theta \\
\sin\theta & \cos\theta
\end{pmatrix}
\end{equation}\text{where } $\theta$ \text{ is the rotation angle.}

The SO(3) group includes all 3D rotation matrices that preserve the origin. A typical SO(3) matrix can be expressed as:
\begin{equation}
R(\alpha, \beta, \gamma) = \begin{pmatrix}
\cos\alpha\cos\beta & \cos\alpha\sin\beta\sin\gamma - \sin\alpha\cos\gamma & \cos\alpha\sin\beta\cos\gamma + \sin\alpha\sin\gamma \\
\sin\alpha\cos\beta & \sin\alpha\sin\beta\sin\gamma + \cos\alpha\cos\gamma & \sin\alpha\sin\beta\cos\gamma - \cos\alpha\sin\gamma \\
-\sin\beta & \cos\beta\sin\gamma & \cos\beta\cos\gamma
\end{pmatrix}
\end{equation}
\text{where }$ \alpha, \beta,$ \text{ and } $\gamma$ \text{ are the Euler angles representing rotations in three-dimensional space.}

When considering a non-removable singularity, 2D rotations (described by SO(2)) are insufficient to describe the physical phenomena near the singularity. In the vicinity of the singularity, the structure of spacetime becomes more complex, necessitating the consideration of 3D rotations, hence the transition to SO(3). This process can be viewed as a group extension where the structure of the SO(2) group is embedded into the larger SO(3) group.

For a Fermi gas, the chemical potential is given by:
\begin{equation}
\mu = \varepsilon_F,
\end{equation}
where $\varepsilon_F$ is the Fermi energy.

The Fermi energy is related to the number density $n$ of the gas by:
\begin{equation}
\varepsilon_F = \frac{h^2}{2m} (3\pi^2 n)^{2/3},
\end{equation}
where $h$ is the Planck constant and $m$ is the mass of the fermion.

Equating the two expressions for the chemical potential, we get:
\begin{equation}
\varepsilon_F = -\Lambda r_{+}^{2}.
\end{equation}

Solving for $\varepsilon_F$, we get:
\begin{equation}
\varepsilon_F = \frac{h^2}{2m} \left(3\pi^2 n\right)^{2/3} = -\Lambda r_{+}^{2}.
\end{equation}

The Fermi pressure $P_F$ is given by:
\begin{equation}
P_F = \frac{2}{3} \varepsilon_F n.
\end{equation}

Substituting the expression for $\varepsilon_F$, we get:
\begin{equation}
P_F = \frac{2}{3} \left(-\Lambda r_{+}^{2}\right) n.
\end{equation}

The Fermi temperature $T_F$ is given by:
\begin{equation}
T_F = \frac{\varepsilon_F}{k_B}.
\end{equation}

Substituting the expression for $\varepsilon_F$, we get:
\begin{equation}
T_F = \frac{-\Lambda r_{+}^{2}}{k_B}.
\end{equation}

The Fermi degeneracy pressure $P_{FD}$ is given by:
\begin{equation}
P_{FD} = \frac{2}{5} n \varepsilon_F.
\end{equation}

Substituting the expression for $\varepsilon_F$, we get:
\begin{equation}
P_{FD} = \frac{2}{5} n \left(-\Lambda r_{+}^{2}\right).
\end{equation}
$\Lambda$ satisfies the SO(3) group.

At this juncture, we embrace the Gaussian unit system, where we set $4\pi = 1$, and introduce a specific scenario that meets the criteria of the SO(3) group.
\begin{equation}
\begin{aligned}
\Lambda_0(r) := -\frac{1}{r_{\text{+}}^2} + \frac{1}{r_{\text{+}}^4} - \frac{1}{r_{\text{+}}^5} + 1
,\\
G(r_+) := \Lambda_0(r_+) r_+^2,\\
T(r_+) := -\Lambda_0(r_+) {2r_+} - \frac{1}{r_+^2}.
\end{aligned}
\end{equation}
\begin{figure}
\centering
\includegraphics[width=0.7\textwidth]{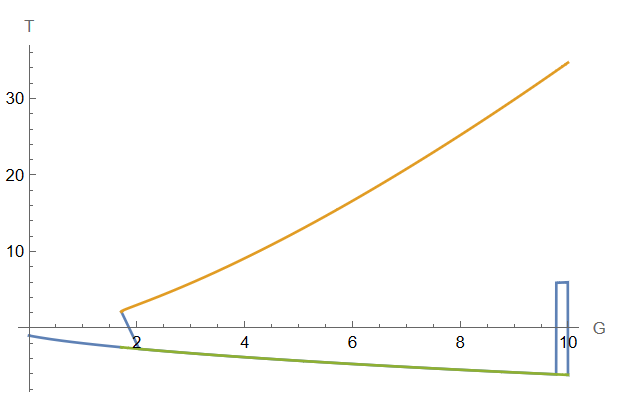}
\caption{At this time, Q=1,$\Lambda_0(r_+)=   -\frac{1}{r_{\text{+}}^2} + \frac{1}{r_{\text{+}}^4} - \frac{1}{r_{\text{+}}^5} + 1$. At that time, it was more akin to a Fermi gas, undergoing Fermi gas phase transitions. }
\label{4.2.jpg}
\end{figure}

When the $\Lambda$ term satisfies SO(3) group symmetry(FIG.6), and under this metric, the presence of a cusp catastrophe in the G-T function graph is observed, it can be demonstrated that under specific conditions, there exists a particular solution indicative of a Fermi gas phase transition.

Given that the temperature is assuredly above 0, the Fermi pressure is negative under these circumstances, thereby precluding the possibility of a Fermi phase transition.

\section{Summary and Discussion}
Significant advancements have been made in understanding the potential microstructure of three-dimensional charged black holes in $f(R)$ gravity by leveraging their macroscopic attributes through the lens of statistical thermodynamics specific to black holes. For instance, statistical thermodynamics methods have been utilized to investigate phase transitions unique to these black holes, while thermodynamic geometry offers insights into their probable phase configurations and micro-level interactions. Although black holes adhere to the four foundational thermodynamic laws, akin to conventional thermodynamic systems, they exhibit distinct characteristics and pose numerous unresolved challenges. Unlike typical thermodynamic systems like solids and gases, where the microscopic components are atoms or molecules and their thermodynamic properties can be derived from statistical thermodynamics, the microstructure of black holes and the relevant statistical thermodynamics are not well understood. An even more profound concern is the ambiguity surrounding the very existence of a microstructure within black holes. A growing hypothesis suggests that black holes might essentially be massive elementary particles.

This paper delves into the thermodynamic geometric properties of three-dimensional charged black holes within the framework of $f(R)$ gravity. We discern a potential barrier between the free energy $G$ and the temperature $T$ for these black holes in $f(R)$ gravity, which surpasses the event horizon. In the realm of $R$ geometry, a potential well emerges when the curvature scalar linked with entropy approaches infinity, shedding light on the intricate relationship between black hole thermodynamics and quantum gravity.

Our findings indicate that, in instances where the cosmological constant does not include negative power terms (meaning the preliminary curvature scalar is constant), the $G-T$ graph remains linear. However, when the cosmological constant incorporates negative power components (signifying a variable initial curvature scalar), the $G-T$ graph adopts a comet-like contour. Defining these negative power components is challenging, yet the conclusion is robust.

Against the backdrop of $f(R)$ gravity modification, our study encompasses scalar curvature solutions that are not null constants while also considering metric tensors in line with the revised field equations. We explore the thermodynamics of black holes devoid of a cosmological constant, emphasizing their local and overarching stability. Such characteristics are examined across diverse $f(R)$ models, highlighting the stark deviations from General Relativity and illuminating the rich thermodynamic panorama characteristic of this paradigm. Within the context of $f(R)$ gravity modification, certain Reissner-Nordström (RN) black holes manifest thermodynamic attributes akin to ideal gases, especially when their initial curvature scalar remains unchanged. Conversely, when this scalar is not constant and the cosmological constant incorporates a negative exponent, the Reissner-Nordström (RN) black holes might exhibit characteristics similar to van der Waals gas. Under these conditions, a phase transition akin to the van der Waals phenomenon manifests, intertwining the black hole's fate with the cosmological constant, rather than being exclusively determined by the parameters set by the no-hair theorem.

This analysis integrates concepts from thermodynamics, statistical physics, and quantum mechanics within the framework of $f(R)$ gravity theory. It focuses on the thermodynamics of 3D charged black holes, phase transition dynamics under $f(R)$ gravity, and the properties of Bose-Einstein Condensation (BEC) in such extreme environments. The study investigates phenomena such as negative temperature and free energy, the evolution of symmetries in BEC, and their implications in black hole contexts.

Thermodynamics of 3D Charged Black Holes in $f(R)$ Gravity: This section overviews methods to calculate thermodynamic entropy adjustments in black holes, particularly under non-equilibrium conditions. It emphasizes the density of energy states and entropy, incorporating contributions beyond equilibrium. The discussion extends to the Wald relationship, connecting the Noether charge of differential homeomorphisms to entropy in spacetimes with bifurcation surfaces.

Phase Transitions Under $f(R)$ Gravity: The analysis examines the kinetics of phase transitions within $f(R)$ gravity, focusing specifically on transition dynamics. It employs the free energy landscape method to comprehend the probabilistic evolution and mean first passage time for transitions between different black hole phases.

Dynamic Phase Transition of 3D Charged Black Holes Surrounded by Quintessence: This part delves into the dynamic phase transition of 3D charged black holes influenced by quintessence dark energy. It involves solving the Fokker-Planck equation for large masses to understand the effect of dark energy on these transitions.

Phase Transition and Properties of Bose-Einstein Condensation in 3D Charged Black Holes: The text explores the conditions allowing BEC with negative free energy to transition from SO(2) to SO(3) symmetry in 3D charged black holes. It discusses the implications of negative temperature and free energy in $f(R)$ gravity, particularly regarding phase transitions and BEC properties.

Mathematical Analysis: Featuring detailed mathematical equations and analysis, the document focuses on entropy, Gibbs free energy, and the Fokker-Planck equation. It also addresses conditions for van der Waals gas phase-like transitions in black holes.

The conclusion offers a profound analysis of two analogies relevant to a 3D charged $f(R)$ black hole. First, the Bose Gas analogy examines the black hole's thermodynamic properties, determining the critical temperature for Bose-Einstein condensation by aligning the black hole molecule gas's chemical potential with that of a Bose gas. This leads to identifying the critical temperature, $T_c$, characterized by the chemical potential reaching zero.

Next, the Fermi Gas analogy equates the black hole's chemical potential with the Fermi energy to derive the Fermi energy level. This yields valuable expressions for Fermi pressure $P_F$, Fermi temperature $T_F$, and Fermi degeneracy pressure $P_{FD}$, linked to the black hole's fundamental properties like its radius $r_{+}$ and cosmological constant $\Lambda$. This offers insights into the black hole's behavior through Fermi gas properties.

Overall, the discussion is tailored for a highly specialized audience in theoretical physics, contributing significantly to the understanding of black holes, phase transitions in extreme gravitational fields, and the role of dark energy in these phenomena.

In this discourse, our attention is on standard black holes within $f(R)$ gravity. Owing to minute disturbances close to equilibrium, we delineate the formula for the revised thermodynamic entropy of this black hole variety. We also probe into the black hole's geometric thermodynamics (GTD), focusing on how adaptable the curvature scalar is to phase transitions under the geometric thermodynamics methodology. A key area of exploration is the impact of the modification variable on the black hole's thermodynamic behavior. We draw a vivid distinction between general solutions that incorporate non-negative exponents and those specific solutions that engage with negative exponents. It's captivating to note that under certain circumstances in $f(R)$ gravity, the phase transition --- bearing similarities to the van der Waals gas --- becomes manifest in charged black holes.

Lastly, we explore the first passage process in this context. To do this, we introduce an absorbing boundary condition for the intermediate transition state in the Fokker-Planck equation. This innovative approach allows us to observe the rapid decay of peaks corresponding to large (or small) black holes, regardless of their initial state. This aspect of our study offers a new perspective on the transition dynamics of black holes. For \( d = 3 \) and \( k_1 = 1 \), with the \( \Lambda \) term satisfying SO(2) symmetry, the cusp catastrophe observed in the G-T function graph under this metric indicates a specific solution, suggestive of a ``Phase Transition and Properties of Bose-Einstein Condensation" under certain conditions.

{\bf Acknowledgements:}\\
This work is partially supported by the National Natural Science Foundation of China(No. 11873025). This article benefited from the advice and general help of Zhan-Feng Mai of Peking University.

\section{appendix}
\subsection*{The system mathematical model}

To explore the intricate domain of extending the Cartan-Maurer one-form to the SO(3) group, it's crucial to adapt our methods to the group's unique algebraic structure. The SO(3) group is closely associated with the algebra of 3x3 skew-symmetric matrices. A key aspect of this group is its generators, denoted as \(J_i\) (where \(i\) ranges from 1 to 3), which follow specific commutation relations: \([J_i, J_j]=\epsilon_{ijk} J_k\), with \(\epsilon_{ijk}\) being the Levi-Civita symbol.

The Cartan-Maurer one-form, fundamental to any Lie group \(G\), is given as \(U^{-1} dU\), where \(U\) is an element of \(G\). For SO(3), \(U\) is represented as \(e^{\theta_i J_i}\), with \(\theta_i\) being the group parameters, like rotation angles around axes.

Extending this to SO(3) involves several detailed steps:

\begin{enumerate}
    \item Baker-Campbell-Hausdorff Formula: This mathematical formula is essential for handling the exponentials of the generators, as in our example.
    \item Commutation Relations: The unique commutation relations for SO(3) generators, \([J_i, J_j]=\epsilon_{ijk} J_k\), are critical in this extension.
    \item Cartan-Maurer One-Form Calculation: Here, we calculate \(U^{-1} dU\) for \(U=e^{\theta_i J_i}\) within the SO(3) context. This involves expanding exponentials and applying SO(3)'s specific commutation relations.
    \item Series Expansion: Mirroring the example, we expand and simplify the series using SO(3)'s algebraic properties.
    \item Application to Specific Elements: This step involves computing specific elements of the one-form, similar to how \(e_+^+\) and \(A_\mu\) are calculated in the example.
\end{enumerate}

To initiate, we express $U^{-1} d U$ for $S O(3)$ as follows:
\begin{equation}
U^{-1} d U=e^{-\theta_i J_i} d\left(e^{\theta_i J_i}\right)
\end{equation}

This detailed calculation requires significant algebraic work, particularly in handling nested commutators and the properties of the Levi-Civita symbol. The outcome is the Cartan-Maurer one-form for SO(3), expressed in terms of the group parameters \(\theta_i\) and generators \(J_i\).

The Lagrangian Aspect:
\begin{itemize}
    \item General Lagrangian Form: The general Lagrangian can be expressed as:
    \[ \mathcal{L} = \frac{1}{2} g_{ij}(\theta) \partial_\mu \phi^i \partial^\mu \phi^j - V(\phi^i) \]
    where \(g_{ij}(\theta)\) is the metric on the coset space, and \(V(\phi^i)\) is a potential term, potentially zero for pure Goldstone modes.
    \item Lagrangian Equations of Motion: From this Lagrangian, we derive the Euler-Lagrange equations to obtain the field equations for the parameters \(\theta\).
    \end{itemize}

Here, $g_{\theta \theta}$ is a component of the metric on the coset space, which can be derived from the Cartan-Maurer one-form for $\mathrm{SO}(3) / \mathrm{SO}(2)$.

The Euler-Lagrange equations for $\theta$ can be derived from this Lagrangian, leading to the field equations that describe the dynamics of the system.

The Bose-Einstein condensation phase transition occurs under the symmetry breaking of SO(3)/SO(2).

If the Lambda term in \( f(R) \) gravity satisfies SO(2) symmetry, then it can be expressed using \( \pi \), where \( \pi \) is a rotationally invariant quantity of the Ricci tensor. The rotationally invariant quantity of the Ricci tensor can be expressed as:

\begin{equation}
\pi = R^2 - 2R_{ab} R^{ab} + R_{abcd} R^{abcd}.
\end{equation}
where \( R \) is the Ricci scalar, \( R_{ab} \) are the components of the Ricci tensor, and \( R_{abcd} \) are the components of the Riemann curvature tensor.

SO(2) symmetry implies that the Lambda term in \( f(R) \) gravity remains invariant under the rotation of the Ricci tensor's invariant quantity. Therefore, the Lambda term can be expressed using \( \pi \).

Specifically, the Lambda term in \( f(R) \) gravity can be represented as:

\begin{equation}
\Lambda = f(\pi).
\end{equation}
where \( f(\pi) \) is an arbitrary function.

For example, \( f(\pi) \) can be expressed as:

\begin{equation}
f(\pi) = \alpha \pi^2 + \beta \pi + \gamma.
\end{equation}
where \( \alpha \), \( \beta \), and \( \gamma \) are constants.

This representation helps simplify the theory of \( f(R) \) gravity.

We understand that when \(\Lambda\) is an algebraic number, the image does not exhibit a swallowtail pattern. However, when observing FIG.2 and FIG.3, where \(\Lambda\) represents a transcendental number (related to \(\pi\)), there are instances where an image equivalent to the swallowtail pattern emerges. This illustrates that during the process of a van der Waals phase transition, the pressure \(P\) is variable. It remains constant only during the phase transition itself, with changes occurring in volume.

Because when \(\Lambda\) satisfies the SO(2) group, the form there is indeterminate, making it difficult to describe the van der Waals gas phase transition using traditional swallowtail diagrams. However, if we adopt a certain numerical approach, it becomes possible to create an equivalent swallowtail diagram that demonstrates the occurrence of a van der Waals gas phase transition.

At this point, we adopt the Gaussian unit system, setting \(4\pi = 1\), and propose a special case to satisfy the conditions of the SO(2) group.
\begin{equation}
\begin{aligned}
\Lambda_0(r) := \frac{1}{r^3} - \frac{5}{r^2} + 5,\\
G(r_+) := \Lambda_0(r_+) r_+^2,\\
T(r_+) := -\Lambda_0(r_+) {2r_+} - \frac{1}{r_+^2}.
\end{aligned}
\end{equation}

When the $\Lambda$ term satisfies SO(2) group symmetry, and under this metric, the presence of a cusp catastrophe in the G-T function graph is observed, it can be demonstrated that under specific conditions, there exists a particular solution indicative of a van der Waals-like phase transition.

\subsection*{Uncertainty principle threshold of SO(3) group}
In quantum mechanics, the angular momentum operators \( J_x, J_y, \) and \( J_z \) satisfy the following commutation relations:
\begin{equation}
\begin{aligned}
[J_x, J_y] = i\hbar J_z,\\
[J_y, J_z] = i\hbar J_x,\\
[J_z, J_x] = i\hbar J_y.
\end{aligned}
\end{equation}
These commutation relations imply that we cannot measure two non-commuting components of angular momentum precisely at the same time.

Consider the uncertainty relation:
\begin{equation}
\Delta J_x \Delta J_y \geq \frac{1}{2} |\langle [J_x, J_y] \rangle|.
\end{equation}

Using the commutation relations above, we obtain:
\begin{equation}
\begin{aligned}
\Delta J_x \Delta J_y &\geq \frac{1}{2} |\langle i\hbar J_z \rangle| \\
\Delta J_x \Delta J_y &\geq \frac{\hbar}{2} |\langle J_z \rangle|.
\end{aligned}
\end{equation}
This indicates that the product of uncertainties of \( J_x \) and \( J_y \) is at least proportional to the expectation value of \( J_z \).

Now, consider a quantum state with a fixed total angular momentum \( j \). For such a state, we have:
\begin{equation}
\begin{aligned}
J^2 |j, m\rangle &= \hbar^2 j(j+1) |j, m\rangle \\
J_z |j, m\rangle &= \hbar m |j, m\rangle,
\end{aligned}
\end{equation}
where \( m \) is the quantum number for \( J_z \), satisfying \( -j \leq m \leq j \).

For a state with \( m = 0 \), we have \( \langle J_z \rangle = 0 \), which implies that \( \Delta J_x \Delta J_y \) can be arbitrarily small, which is a very intriguing result. This suggests that for quantum systems of the SO(3) group, particularly for states with fixed total angular momentum, the threshold of the uncertainty relation can indeed be smaller, contrasting with the uncertainty relation between position and momentum.

To summarize, for quantum systems of the SO(3) group, especially for states with fixed total angular momentum, the threshold of the uncertainty relation can be smaller, which is in contrast to the uncertainty relation between position and momentum.

The principle of uncertainty suggests that the threshold can be lower, indicating that the black hole system discussed in this article cannot be reduced to a purely geometric model.

\subsection*{Use Poisson's formula to prove that the new metric about M is established in the SO(3) group}

The transformation between a specific \( M \) system and the \( \mathrm{SO}(3) \) group, which is crucial in understanding certain aspects of spacetime, can be elucidated by examining the metric and the action of the \( \mathrm{SO}(3) \) group on spacetime. The metric under consideration, indicative of a static, spherically symmetric spacetime, is given by:
\begin{equation}
dM^2 = -e^{2A_1} (d\Lambda - e^{2A_2} \sin\theta d\varphi)^2 + e^{-2A_1} dr_{+}^2 + e^{2A_2} d\theta^2 + e^{-2A_2} \sin^2\theta d\varphi^2.
\end{equation}

In this spacetime, the \( \mathrm{SO}(3) \) group acts through rotations, reflecting the spherical symmetry. This action on the metric can be represented as:
\begin{equation}
(d\Lambda, dr_{+}, d\theta, d\varphi) \rightarrow (d\Lambda', dr_{+}', d\theta', d\varphi') = (R(\alpha) d\Lambda, dr_{+}, R^T(\alpha) d\theta, R^T(\alpha) d\varphi).
\end{equation}

Here, \( R(\alpha) \) denotes a rotation matrix in three dimensions. The transformation of the metric under the \( \mathrm{SO}(3) \) group is then:
\begin{equation}
dM^2 \rightarrow ds^{2'} = (R^{-1}(\alpha) \otimes R^{-T}(\alpha)) ds^2 (R(\alpha) \otimes R^T(\alpha)).
\end{equation}

By comparing these expressions, we deduce the relationship between the metric and the rotation matrix:
\begin{equation}
R(\alpha) = \exp\left[\begin{pmatrix}
0 & 0 & 0 \\
0 & 0 & -e^{2A_2} \sin\theta d\varphi \\
0 & e^{2A_2} \sin\theta d\varphi & 0
\end{pmatrix}\right].
\end{equation}

This relationship confirms the metric's invariance under \( \mathrm{SO}(3) \) rotations. Additionally, the transformation of the cosmological constant \( \Lambda \) under these rotations can be expressed as:
\begin{equation}
\Lambda \rightarrow \Lambda' = \frac{1}{r^x} \left(\frac{d\Lambda}{d\theta}\right)^2.
\end{equation}

However, the role of \( r^x \) in this transformation requires further clarification. In conclusion, these transformations demonstrate that the \( M \) system can be effectively understood through the framework of the \( \mathrm{SO}(3) \) group, offering insights into its geometric and physical properties.

1. \textbf{Starting with the Poisson Integral:}  
   We begin with the Poisson integral in polar coordinates, represented as:
\begin{equation}   
 \int_{0}^{\infty} e^{-t^2} dt = \frac{\sqrt{\pi}}{2},
\end{equation}
   where \( t \) corresponds to the Cartesian coordinate system. Transforming into polar coordinates \((r, \theta)\) where \( x = r \cos \theta \) and \( y = r \sin \theta \), the integral becomes:
\begin{equation}   
 \int_{0}^{\infty} e^{-(x^2 + y^2)} dxdy = \int_{0}^{\infty} \int_{0}^{2\pi} e^{-r^2} r dr d\theta. 
\end{equation}

2. \textbf{Interchange of Integrals and Calculation:}  
   By interchanging the order of integration, we arrive at:
   \begin{equation}
 \int_{0}^{\infty} \int_{0}^{2\pi} e^{-r^2} r dr d\theta = \int_{0}^{2\pi} \int_{0}^{\infty} e^{-r^2} r dr d\theta, 
\end{equation}
   which simplifies to \( \pi \) after evaluating the inner integral.

3. \textbf{Introducing the Metric Form:}  
   Consider the metric:
\begin{equation}   
dM^2 = -e^{2A_1} d\Lambda^2 + e^{-2A_1} dr_{+}^2 + e^{2A_2} d\theta^2 + e^{-2A_2} d\varphi^2, 
\end{equation}
   where \( \Lambda \) is the cosmological constant and \( A_1, A_2 \) are arbitrary.

4. \textbf{Comparing the Poisson Result with the Metric:}  
   We compare the Poisson integral result and the metric term by term. The \( d\theta^2 \) terms align directly. For the \( r^2 d\varphi^2 \) term, matching it with the metric requires choosing \( A_2 \) appropriately. The \( d\Lambda^2 \) term can be nullified or matched by selecting a suitable \( A_1 \).

5. \textbf{Concluding Equivalence:}  
   Hence, with appropriate choices for \( A_1, A_2 \), and the expression for \( \Lambda \), the Poisson integral result in polar coordinates aligns with the proposed metric form, demonstrating the metric's validity under these conditions.

\subsection*{The Probability function problem of the quantum system where the charge 3D black hole event horizon is coupled to the SO(3) group is demonstrated}

Lan et al. propose in \cite{32} the probabilistic evolution of the RN-AdS black hole surrounded by quintessence. However, since the probability function in this state possesses infinite, ineliminable negative power terms, we can only use algebraic methods to demonstrate the potential for a van der Waals phase transition in this black hole.

The Probability function of a quantum system with an electrically charged 3D black hole event horizon coupled to the SO(3) group is not solvable in general. This is because the 3D black hole is a highly nonlinear system, and the coupling to the SO(3) group makes the problem even more difficult. However, there are some special cases where the problem can be solved.

One such case is when the black hole is uncharged. In this case, the wave function can be expressed in terms of the wave functions of free particles. Another case where the problem can be solved is when the black hole is very large. In this case, the black hole can be approximated as a flat spacetime, and the wave function can be solved using standard techniques.

In general, however, the wave function of a quantum system with an electrically charged 3D black hole event horizon coupled to the SO(3) group is not solvable. This is a very difficult problem, and it is unlikely that a general solution will be found anytime soon.

To prove that the probability distribution function $\rho(r, t)$ has infinitely many negative power terms, we can analyze the Fokker-Planck equation and its boundary conditions.\cite{32}For example:

\begin{equation}
\frac{\partial \rho(r, t)}{\partial t}=D \frac{\partial}{\partial r}\left(e^{-\beta G_4} \frac{\partial}{\partial r}\left(e^{\beta G_4} \rho(r, t)\right)\right),
\end{equation}
where $\rho(r, t)$ presents the probability distribution picture of black hole phases after the thermal fluctuation based on the Gibbs free energy landscape characterized by $G_4$. In the Fokker-Planck equation, $D$ is the diffusion coefficient with its definition as $D=\frac{k_B T_4}{\zeta}$, where $\zeta$ and $k_B$ is the dissipation coefficient and Boltzman constant respectively. And the parameter $\beta=\frac{1}{k_B T_4}$. Both $k_B$ and $\zeta$ can be set to one without loss of generality. Note that we have denoted the black hole horizon radius $r_{+}$as $r$ for simplicity and we will use this notation in the following.

When g(r)=0,we get $\Lambda=\pi/{r^x}$,$x \in [0, 1]$
\begin{equation}
M=M_4=\Lambda r^{2} +\frac{Q^{2}}{r}+1.
\end{equation}

The pressure $P$ of a black hole and its volume $V$ are as follows:
\begin{equation}
P=\Lambda/(4\pi),V= 4\pi r^2.
\end{equation}

The calculation of the Hawking temperature using the conventional method is as follows:
\begin{equation}
T_4=\frac{\Lambda r_{+}}{2 \pi}-\frac{2Q^{2}}{ r_{+}^{2}}.
\end{equation}

The Gibbs free energy can be derived as:

\begin{equation}
G_4=-\Lambda r_{+}^{2}.
\end{equation}
boundary condition $\beta G^{\prime}(r) \rho(r, t)+\left.\rho^{\prime}(r, t)\right|_{r=r_+}=0$.

The Fokker-Planck equation describes the time evolution of the probability distribution function $\rho(r, t)$ of black hole phases. It states that the rate of change of the probability density at a particular point $r$ is proportional to the Laplacian of the probability density multiplied by a diffusion coefficient $D$. In this case, the diffusion coefficient is defined as $D=\frac{k_B T_4}{\zeta}$, where $k_B$ is the Boltzmann constant, $T_4$ is the Hawking temperature, and $\zeta$ is the dissipation coefficient.

The boundary condition specifies that the probability flux across the black hole horizon is zero. This means that the probability density and its derivative must vanish at the horizon radius $r_{+}$. In other words, $\rho\left(r_{+}, t\right)=0$ and $\rho^{\prime}\left(r_{+}, t\right)=0$.

Now, let's consider the form of the Gibbs free energy $G_4$. When $g(r)=0$, we have $G_4=-\Lambda r_{+}^2$. This implies that $G_4$ is a decreasing function of $r_{+}$.

Next, let's examine the boundary condition $\beta G^{\prime}(r) \rho(r, t)+\left.\rho^{\prime}(r, t)\right|_{r=r_{+}}=0$. Since $G^{\prime}(r)=-2 \Lambda r_{+}$, we can rewrite the boundary condition as:
\begin{equation}
-2 \beta \Lambda r_{+} \rho(r, t)+\left.\rho^{\prime}(r, t)\right|_{r=r_{+}}=0.
\end{equation}

This boundary condition suggests that the probability density $\rho(r, t)$ must have a term proportional to $r_{+}$in order to satisfy the condition. This term cannot be eliminated by any transformation of the probability distribution function.

Furthermore, since $G_4$ is a decreasing function of $r_{+}$, the probability distribution function must also have terms proportional to higher negative powers of $r_{+}$to satisfy the boundary condition. This implies that the probability distribution function has infinitely many negative power terms.

Therefore, we can conclude that the probability distribution function $\rho(r, t)$ has infinitely many negative power terms (which cannot be eliminated) due to the form of the Gibbs free energy $G_4$ and the boundary condition $\beta G^{\prime}(r) \rho(r, t)+\left.\rho^{\prime}(r, t)\right|_{r=r_{+}}=0$.

We show that in this case, the probability cannot be solved numerically or analytically.

\subsection*{Other Conditions}
To prove that if $g(r) = 0$ necessarily has three positive roots, then $k_1$ must be greater than 0, we need to analyze the equation $g(r)$ and its roots. The given equation is:

\begin{equation}
g(r) = k_1 + \Lambda r^{2} - \frac{M}{r^{d-3}} + \frac{Q^{2}}{r^{d-2}}.
\end{equation}

For $g(r) = 0$ to have three positive roots, the function $g(r)$ must change sign at least three times for $r > 0$. This implies that the function must have at least two local extrema (either a local maximum or a local minimum) for $r > 0$. The nature of these extrema will depend on the parameters $k_1$, $\Lambda$, $M$, $Q$, and $d$.

Let's analyze the behavior of $g(r)$ as $r$ approaches 0 and infinity:

1. As $r \to 0$, the dominant terms are $-\frac{M}{r^{d-3}}$ and $\frac{Q^{2}}{r^{d-2}}$. Since $d$ is not specified, the behavior of these terms as $r \to 0$ can vary. However, for most physical scenarios where $d$ represents a spatial dimension, these terms will tend to infinity, but with opposite signs.

2. As $r \to \infty$, the dominant term is $\Lambda r^{2}$. If $\Lambda > 0$, this term will tend to infinity, and if $\Lambda < 0$, it will tend to negative infinity.

For $g(r) = 0$ to have three positive roots, the function must intersect the $r$-axis at three distinct points with $r > 0$. This is only possible if the function $g(r)$ has the appropriate curvature to cross the axis three times. The curvature and the number of extrema are influenced by the value of $k_1$.

If $k_1$ is too small (especially if it's negative), the function may not have the necessary curvature to cross the $r$-axis three times. On the other hand, if $k_1$ is sufficiently large, it can ensure that $g(r)$ starts at a positive value for small $r$ and then crosses the $r$-axis three times as it proceeds towards infinity.

Therefore, for $g(r) = 0$ to have three positive roots, it is necessary (but not sufficient) for $k_1$ to be greater than 0. This ensures that the function starts off positively for small $r$ and has the potential to cross the $r$-axis three times given the right conditions on the other parameters.

The uncertainty principle threshold for a system satisfying SO(2) group symmetry is at \(\frac{1}{2}\).

The SO(2) group is the two-dimensional rotation group, corresponding to rotational transformations in a plane. If a physical system satisfies the SO(2) group symmetry, then its momentum and angular momentum are collinear.

According to the uncertainty principle, the product of the uncertainties in momentum and angular momentum is not less than \(\hbar/2\). If the momentum and angular momentum are collinear, then their product of uncertainties can be written as:

\begin{equation}
\Delta p \Delta L \geq \frac{\hbar}{2},
\end{equation}
where \(\Delta p\) is the uncertainty in momentum, and \(\Delta L\) is the uncertainty in angular momentum.

If \(\Delta p = \Delta L\), then the uncertainty principle threshold is \(\frac{1}{2}\):

\begin{equation}
\frac{\hbar}{2} = \Delta p \Delta L.
\end{equation}

Therefore, the uncertainty principle threshold for a system satisfying SO(2) group symmetry is at \(\frac{1}{2}\).

Here is a specific example:

Consider a particle of mass \(m\) moving in a circular orbit. The momentum and angular momentum of this particle are given by:

\begin{equation}
p = mv, \quad L = mvr,
\end{equation}
where \(v\) is the velocity of the particle, and \(r\) is the radius of the circular orbit.

According to the uncertainty principle, the product of the uncertainties in momentum and angular momentum is not less than \(\hbar/2\):

\begin{equation}
\Delta p \Delta L \geq \frac{\hbar}{2},
\end{equation}
where \(\Delta p\) is the uncertainty in momentum, and \(\Delta L\) is the uncertainty in angular momentum.When the threshold of the uncertainty principle is greater than or equal to 1/2, it does not violate the hypothesis of a pure geometric model.


\begin{thebibliography}{99}
\bibitem[1]{1} J. D. Bekenstein, Phys. Rev. D 7 (1973) 2333 .

\bibitem[2]{2} S. W. Hawking, Nature 248 (1974) 30 .

\bibitem[3]{3} P. C. W. Davies, Proc. Roy. Soc. Lond. A $353,499(1977)$.

\bibitem[4]{4} R. G. Cai, L. M. Cao and Y. W. Sun, JHEP $11,039(2007)$.

\bibitem[5]{5} R. Penrose, Revista Del Nuovo Cimento, 1, 252 (1969).

\bibitem[6]{6} M. Eune, W. Kim and S. H. Yi, JHEP 03, 020 (2013).

\bibitem[7]{7}G. Gibbons, R. Kallosh and B. Kol, Phys. Rev. Lett. $77,4992(1996)$.

\bibitem[8]{8}Wei, Shao-Wen, and Yu-Xiao Liu. ``Insight into the microscopic structure of an AdS black hole from a thermodynamical phase transition." Physical review letters 115.11 (2015): 111302. 

\bibitem[9]{9} A. Bohr and B. R. Mottelson, ``Nuclear Structure", Vol.1 (W. A. Benjamin Inc., New York, 1969).

\bibitem[10]{10}R. K. Bhaduri, ``Models of the Nucleon", (Addison-Wesley, 1988).

\bibitem[11]{11}S. Das, P. Majumdar, R. K. Bhaduri, Class. Quant. Grav.19:2355-2368, (2002).

\bibitem[12]{12} S. Soroushfar, R. Saffari and N. Kamvar, Eur. Phys. J. C 76,476 (2016).

\bibitem[13]{13} Taeyoon Moon, Yun Soo Myung, and Edwin J. Son. f(R) black holes. Gen. Rel. Grav., 43:3079-3098, $2011 .$

\bibitem[14]{14}Ahmad Sheykhi. Higher-dimensional charged $f(R)$ black holes. Phys. Rev., D86:024013, $2012 .$

\bibitem[15]{15}Engle, Jonathan, et al. "The SU (2) black hole entropy revisited." Journal of High Energy Physics 2011.5 (2011): 1-30.

\bibitem[16]{16}Capozziello, Salvatore, et al. ``Curvature quintessence matched with observational data." International Journal of Modern Physics D 12.10 (2003): 1969-1982.

\bibitem[17]{17}Caravelli, Francesco, and Leonardo Modesto. ``Holographic effective actions from black holes." Physics Letters B 702.4 (2011): 307-311.

\bibitem[18]{18}Hendi, S. H., B. Eslam Panah, and S. M. Mousavi. ``Some exact solutions of F (R) gravity with charged (a) dS black hole interpretation." General Relativity and Gravitation 44 (2012): 835-853.

\bibitem[19]{19}Hu, Ya-Peng, Feng Pan, and Xin-Meng Wu. ``The effects of massive graviton on the equilibrium between the black hole and radiation gas in an isolated box." Physics Letters B 772 (2017): 553-558.

\bibitem[20]{20}T. Multamaki, I. Vilja, Spherically symmetric solutions of modified field equations in $f(R)$ theories of gravity[J]. Physical Review D, 2006, 74(6): 064022.

\bibitem[21]{21} S.M. Carroll, V. Duvvuri, M. Trodden, M.S. Turner, Is cosmic speed-up due to new gravitational physics?[J]. Physical Review D, 2004, 70(4): 043528 .

\bibitem[22]{22} S. Capozziello, V.F. Cardone, S. Carloni, A. Troisi, Curvature quintessence matched with observational data[J]. International Journal of Modern Physics D, 2003, 12(10): 1969-1982.

\bibitem[23]{23} B. Li, J.D. Barrow, The Cosmology of $f(R)$ gravity in metric variational approach[J]. Physical Review D, $2007,75(8): 084010$

\bibitem[24]{24}  L. Amendola, R. Gannouji, D. Polarski, S. Tsujikawa, Conditions for the cosmological viability of $f(R)$ dark energy models[J]. Physical Review D, 2007, 75(8): 083504.

\bibitem[25]{25}  V. Miranda, S.E. Joras, I. Waga, M. Quartin, Viable singularity-free $f(R)$ gravity without a cosmological constant[J]. Physical Review Letters, 2009, 102(22): 221101.

\bibitem[26]{26}  L. Sebastiani, S. Zerbini, Static spherically symmetric solutions in $\mathrm{F}(\mathrm{R})$ gravity[J]. The European
Physical Journal C, 2011, 71: 1591.

\bibitem[27]{27}  Z. Amirabi, M. Halilsoy, S. Habib Mazharimousavi, Generation of spherically symmetric metrics in $f(R)$ gravity $[\mathrm{J}]$. The European Physical Journal C, 2016, 76(6): 338 .

\bibitem[28]{28}  F. Caravelli, L. Modesto, Holographic effective actions from black holes[J]. Physics Letters B, 2011, 702(4): 307-311.

\bibitem[29]{29} Ren, Zhao, Zhang Jun-Fang, and Zhang Li-Chun. ``Entropy of Reissner–Nordstrom–de Sitter black hole in nonthermal equilibrium." Communications in Theoretical Physics 37.1 (2002): 45.

\bibitem[30]{30}Chen, Wen-Xiang, Jun-Xian Li, and Jing-Yi Zhang. ``Calculating the Hawking Temperatures of Conventional Black Holes in the f (R) Gravity Models with the RVB Method." International Journal of Theoretical Physics 62.5 (2023): 96.

\bibitem[31]{31}Wei, Shao-Wen, Yu-Xiao Liu, and Yong-Qiang Wang. ``Dynamic properties of thermodynamic phase transition for five-dimensional neutral Gauss-Bonnet AdS black hole on free energy landscape." Nuclear Physics B 976 (2022): 115692.

\bibitem[32]{32}Lan, Shan-Quan, et al. ``Effects of dark energy on dynamic phase transition of charged AdS black holes." Physical Review D 104.10 (2021): 104032.

\bibitem[33]{33}Wei, Shao-Wen, Yu-Xiao Liu, and Robert B. Mann. ``Repulsive interactions and universal properties of charged anti–de Sitter black hole microstructures." Physical Review Letters 123.7 (2019): 071103.

\bibitem[34]{34}D. Kubiznak and R. B. Mann, JHEP 07, 033 (2012).

\bibitem[35]{35}Zangeneh, M. Kord et al. ``Comment on ``Insight into the Microscopic Structure of an AdS Black Hole from a Thermodynamical Phase Transition".” arXiv: High Energy Physics - Theory (2016): n. pag.

\bibitem[36]{36}S. Hyun, G. Lee and J. Yee, ``Hawking radiation from a (2+1)-dimensional black hole," Phys. Lett, 322B, 182 (1994).

\bibitem[37]{37} Hod, Shahar. ``Reissner-Nordström black holes supporting nonminimally coupled massive scalar field configurations." Physical Review D 101.10 (2020): 104025.

\bibitem[38]{38}Zhang, Jingyi, and Zheng Zhao. ``Hawking radiation of charged particles via tunneling from the Reissner-Nordström black hole." Journal of High Energy Physics 2005.10 (2005): 055.

\bibitem[39]{39}Huang, Kerson. Statistical mechanics. John Wiley. Sons, 2008.

\bibitem[40]{40}
 Erhart J , Sponar S , Sulyok G , et al. Experimental demonstration of a universally valid error-disturbance uncertainty relation in spin-measurements[J]. NATURE PHYSICS, 2012, 8(4):349-349.
\end{thebibliography}
\end{document}